\documentclass[apj,numberedappendix]{emulateapj}
\usepackage{epsfig}
\usepackage{amsmath}
\usepackage{enumerate}
\usepackage{natbib}
\usepackage{hyperref}
\def\sT{\sigma_{\rm T}}
\def\beq{\begin{equation}}
\def\eeq{\end{equation}}

\def\Rph{R_\star}

\def\Sect{Section}
\def\Sects{Sections}
\def\Eq{Equation}
\def\Eqs{Equations}

\def\Urad{U_\gamma}

\def\dn{\dot{n}}

\def\tIC{t_{\rm IC}}

\def\TC{T_{\rm C}}
\def\tC{t_{\rm IC}}

\def\fKN{f_{\rm KN}}

\def\tsh{t_c}
\def\D{L}
\def\Prad{P_{\rm rad}}
\def\Urad{U_{\rm rad}}

\def\vp{v_{\rm p}}

\def\B{{\cal B}}

\def\Ppl{P_{\rm pl}}

\def\pmax{p_{\max}}
\def\eps{\epsilon_{\rm rad}}
\def\epsB{\epsilon_B}

\def\tauT{\tau_{\rm T}}

\def\rhof{\tilde{\rho}}
\def\adind{\alpha}
\def\vp{v_0^\prime}
\def\xx{x_0^\star}
\def\xs{x_\star}
\def\ts{t_\star}
\def\td{t_{\rm dec}}
\def\xd{x_{\rm dec}}
\def\rhod{\rho_{\rm dec}}
\def\rhob{\rho_{\rm bal}}

\def\taus{\tau_\star}
\def\lsh{l_{\rm sh}}
\def\taush{\Delta\tau}

\def\lph{l_{\rm ph}}
\def\lCoul{l_{\rm Coul}}

\def\tCoul{t_{\rm Coul}}

\def\Uu{U_u}
\def\Ud{U_d}
\def\Pu{P_u}
\def\gu{\gamma_u}
\def\bu{\beta_u}

\def\rhofu{\tilde{\rho}_u}
\def\rhofd{\tilde{\rho}_d}
\def\Pd{P_d}
\def\gd{\gamma_d}
\def\bd{\beta_d}

\def\If{\tilde{I}}
\def\lln{\ell_n}
\def\lph{\ell_{\rm ph}}
\def\a{\sigma}
 \def\wdo{w_d}
 \def\wu{w_u}
 \def\sigd{\sigma_d}
 \def\sigu{\sigma_u}
 \def\pd{p_d}
 \def\pu{p_u}
 \def\vd{v_d}

\def\M{{\cal M}}
\def\Mu{{\cal M}_u}
\def\Md{{\cal M}_d}

\def\bE{{\mathbf E}}
\def\bB{{\mathbf B}}
\def\bv{{\mathbf v}}

\def\sigcr{\sigma_\star}
\def\Heff{H_{\rm eff}}
\def\Peff{P_{\rm eff}}

\def\bth{\beta_{\rm th}}
\def\gth{\gamma_{\rm th}}
\def\fKN{f_{\rm KN}}
\def\tIC{t_{\rm IC}}

\def\lgg{\ell_{\gamma\gamma}}
\def\dnann{\dot{n}_{\rm ann}}
\def\dngg{\dot{n}_{\gamma\gamma}}
\def\sgg{\sigma_{\gamma\gamma}}
\def\nMeV{n_{\rm MeV}}
\def\EIC{E_{\rm IC}}

\def\lIC{\ell_{\rm IC}}
\def\gcr{\gamma_{\rm cr}}
\def\MMeV{{\cal M}_{\rm MeV}}
\def\epsp{\epsilon_p}
\def\eps{\epsilon}
\def\tann{t_{\rm ann}}
\def\gthe{\gamma_{\rm th,e}}

\newbox\grsign \setbox\grsign=\hbox{$>$} \newdimen\grdimen \grdimen=\ht\grsign
\newbox\simlessbox \newbox\simgreatbox \newbox\simpropbox
\setbox\simgreatbox=\hbox{\raise.5ex\hbox{$>$}\llap
     {\lower.5ex\hbox{$\sim$}}}\ht1=\grdimen\dp1=0pt
\setbox\simlessbox=\hbox{\raise.5ex\hbox{$<$}\llap
     {\lower.5ex\hbox{$\sim$}}}\ht2=\grdimen\dp2=0pt
\setbox\simpropbox=\hbox{\raise.5ex\hbox{$\propto$}\llap
     {\lower.5ex\hbox{$\sim$}}}\ht2=\grdimen\dp2=0pt
\def\simgt{\mathrel{\copy\simgreatbox}}
\def\simlt{\mathrel{\copy\simlessbox}}

\shorttitle{Sub-photospheric shocks in relativistic explosions}
\shortauthors{A. M. Beloborodov}

\begin{document}

\title{Sub-photospheric shocks in relativistic explosions}
\author{Andrei M. Beloborodov}
\affil{Physics Department and Columbia Astrophysics Laboratory, Columbia University, 538 West 120th Street, New York, NY 10027, USA
}

\label{firstpage}
\begin{abstract}
This paper examines the mechanism of internal shocks in opaque relativistic outflows, 
in particular in cosmological gamma-ray bursts. The shocks produce neutrino emission 
and affect the observed photospheric radiation from the explosion. They develop from 
internal compressive waves and can be of different types depending on the composition 
of the outflow: 
(1) Shocks in ``photon gas,'' with negligible plasma inertia, have 
a unique structure determined by the force-free condition---zero radiation flux 
in the plasma rest frame. Radiation dominance over plasma inertia suppresses
formation of collisionless shocks mediated by collective electromagnetic fields.
(2) If the outflow is sufficiently magnetized, a strong collisionless subshock develops,
which is embedded in a thicker radiation-mediated structure. 
(3) Waves in outflows with a free neutron component lead to dissipation 
through nuclear collisions.
At large optical depths, shocks have a thickness comparable to the neutron 
free path, with an embedded radiation-mediated and collisionless subshocks.
The paper also presents first-principle simulations of magnetized flows filled 
with photons, demonstrating formation of shocks and their structure.
Simple estimates show that magnetized sub-photospheric shocks are efficient 
producers of photons and have a great impact on the observed photospheric radiation.
The shock structure changes as the outflow expands toward its photosphere.
The dissipation is accompanied by strong $e^\pm$ pair creation,
and the $e^\pm$-dressed shock carries the photosphere with it up to two 
decades in radius, emitting a strong pulse of nonthermal radiation.
\end{abstract}


%############################################################

\section{Introduction}

Astrophysical explosions and jets generate shock waves, which produce radiation. 
Their radiative properties are determined by the dissipation mechanism that 
sustains the velocity jump in the shock and by its ability to generate 
nonthermal particles. This paper examines the 
mechanism of internal shocks in gamma-ray bursts (GRBs) that occur 
before the GRB jets become transparent to radiation. The approach and some of the 
results may also be of interest for other explosions, e.g. in novae or supernovae.

\subsection{Internal shocks in GRB jets}

The main features of GRB explosions may be summarized as follows:
the outflow is relativistic, it carries magnetic fields frozen in fully ionized
plasma, and a large fraction of its energy is carried by neutral particles --- 
photons and free neutrons. 
GRB outflows start very opaque near the central engine of the explosion and 
become transparent at a large ``photospheric'' radius $\Rph$. 
Internal shocks can develop below and above the photosphere. 

Early works proposing internal shocks in GRBs focused on shocks above the photosphere 
(Rees \& M\'esz\'aros 1994; Kobayashi et al. 1997; Daigne \& Mochkovitch 1998).
They can only be collisionless, i.e. mediated by collective 
electromagnetic fields. Their mechanism has been studied in detail 
using particle-in-cell simulations, and it was found that the presence of transverse 
magnetic fields renders the shock unable to accelerate particles 
(Sironi \& Spitkovsky 2011): the postshock electron-ion plasma is in a two-temperature 
state, $T_e<T_i$, with electrons and ions forming nearly Maxwellian distributions.
This may, however, change if the electron-ion outflow is loaded with
$e^\pm$ plasma. Ultra-relativistic shocks in $e^\pm$-loaded plasma with transverse
magnetic field were found capable of accelerating positrons 
(Hoshino et al. 1992; Amato \& Arons 2006; Stockem et al. 2012).

Shocks in opaque plasma below the photosphere $\Rph$ are less explored
and may be key to understanding GRB emission (M\'esz\'aros \& Rees 2000b;
Pe'er et al. 2006; Beloborodov 2010; Levinson 2012).
There is significant evidence that GRB radiation is mainly produced
below the photosphere (Ryde et al. 2011; Beloborodov 2013; Yu et al. 2015),
and detailed simulations of radiative transfer in opaque heated jets 
give spectra consistent with GRB observations (Vurm \& Beloborodov 2016). 
Internal shocks provide a plausible mechanism for sub-photospheric heating
invoked by these models.

Internal shocks may result from the fast variability of the central engine or the 
outflow interaction with the progenitor star (e.g. Lazzati et al. 2013; Ito et al. 2015).
At later stages of ballistic expansion with a high Lorentz factor $\Gamma$, 
internal shocks at radius $r$ can develop from velocity variations on 
scale $L\sim r/\Gamma$ (measured in the outflow rest frame).
This scale also sets the characteristic optical depth seen by photons in the expanding 
outflow, $\tauT\sim \sT n_e\D$, where $\sT$ is Thomson cross section and $n_e$ is 
the proper density of electrons and positrons.

The present paper is motivated by the following questions:
\\
(1) Can sub-photospheric shocks be collisionless? This is assumed in models of 
TeV neutrino emission from the jet-progenitor interaction (Razzaque et al. 2003), however 
the assumption is questionable (Murase \& Ioka 2013).
\\
(2) Is the shock capable of producing high-energy particles? The presence of high-energy 
electrons at large optical depths would have a strong effect on the photospheric radiation
(Pe'er \& Waxman 2004; Beloborodov 2010; Vurm \& Beloborodov 2016).
\\
(3) How does the shock evolve as it emerges from the photosphere and what 
is its observational appearance?

\subsection{Radiation-mediated shocks (RMS)}

Since GRB jets carry a large number of photons per electron,
sub-photospheric shocks are naturally expected to be mediated by radiation
(Levinson 2012).
Then dissipation and the profile of the velocity jump are controlled by photon scattering. 

Basic features of radiation-mediated shocks (RMS) were studied in 
the 1950s (see Zeldovich \& Raizer 1966 and refs. therein). 
The RMS propagation is sustained by radiation diffusion:
radiation generated by the shock diffuses
upstream and pre-heats the upstream gas. This creates a pressure gradient, a kind of 
a ``pillow'' that allows the gas to smoothly decelerate, avoiding the 
collapse of the shock thickness to the collisionless scale (the ion Larmor radius).

The first RMS models assumed that radiation is everywhere in local 
thermodynamic equilibrium. This assumption can be strongly violated in 
astrophysical explosions, as the timescale to establish thermodynamic equilibrium 
can be much longer than the time it takes the gas to cross the shock. 
Models relaxing the equilibrium assumption have 
been developed and applied to supernova shock breakout (e.g. Weaver 1976;
Sapir et al. 2013). The RMS model was also extended to relativistic shocks
(Levinson \& Bromberg 2008; Budnik et al. 2010; Bromberg et al. 2011; Levinson 2012).
The highest temperature achieved in the RMS depends on the photon number 
carried by the upstream through the shock. Levinson (2012) 
emphasized the low efficiency of photon production by the RMS in GRB jets and
developed a shock model with a conserved photon number.

The RMS thickness is large, comparable to or larger than the photon mean free path. 
This inhibits diffusive acceleration of charged particles. In particular, electrons radiate
energy faster than they can cross the shock. The RMS is only capable of a slow 
diffusive acceleration of photons up to the MeV band (in the shock frame).

The RMS picture of internal shocks has, however, a few caveats. Previous work did 
not take into account that the outflow is magnetized, and the magnetic field can 
change the RMS structure and the dissipation mechanism. In addition, GRB explosions 
are expected to carry free neutrons; their collisions can play a key role in shaping the 
shock waves at large optical depths and offer an additional mechanism for producing 
high-energy particles and neutrinos.

\subsection{Outline of the paper}

We begin with basics of shock formation.
Section~2 examines how a super-sonic compressive wave steepens 
and launches a pair of shock waves. We first describe shock formation
in a polytropic gas using the hydrodynamic 
approximation (zero mean free path of all particles and photons). 
Then we relax this approximation and discuss the role of photon 
diffusion in the formation of RMS and collisionless shocks.
We consider a ``cold'' gas with sound speed $c_0\ll c$ and formulate two conditions 
for the immediate RMS formation (vs. formation of a collisionless shock). 
We also discuss flows with large $c_0$,
including the extreme regimes where the flow inertia is 
dominated by radiation ($c_0=c/\sqrt{3}$) or magnetic fields ($c_0=c$).

\Sect~3 describes the general jump conditions for shocks in media with any thermal 
pressure and magnetization. A moderate magnetization of the flow changes 
the jump conditions and we argue that this leads to the formation of a thin collisionless 
subshock, even at large optical depths. We evaluate the region in the parameter space
where a strong collisionless subshock must exist.

\Sect~\ref{sec:photon_gas} focuses on shocks in ``photon-gas'' with sub-dominant 
magnetic fields and negligible plasma inertia ($c_0=c/\sqrt{3}$). This regime may occur 
in GRB explosions at their early stages, during the jet breakout and its acceleration 
by radiation pressure. We use a self-consistent simulation of time-dependent radiative 
transfer and obtain the solution for the shock structure.

Then \Sect~5 presents the RMS structure at later stages when the plasma inertia 
becomes important. As the main tool, we use direct Monte-Carlo simulations of 
time-dependent radiative transfer coupled with the flow dynamics. 
We first investigate RMS formation in a weakly magnetized flow, and discuss the 
effect of bulk Comptonization and $e^\pm$ creation inside the shock front. 
Then we turn to shocks in a magnetized fluid and demonstrate the formation of 
a strong collisionless subshock embedded in the RMS, as anticipated in \Sect~3.

\Sect~6  investigates how plasma heating in a collisionless (sub)shock
results in ``breeding'' of $e^\pm$ pairs.
\Sect~7 discusses the emergence of a $e^\pm$-dressed shock from the photosphere.

\Sect~8 describes shocks in outflows with a free neutron component. 
Due to their large free paths, neutrons introduce a large effective viscosity. 
Nuclear collisions produce ultra-relativistic $e^\pm$ pairs and neutrinos. 
They sustain broad shock fronts until the jet reaches the neutron decoupling radius 
$R_n$ where most neutrons begin to flow freely without collisions.

The results and their implications for GRB models are discussed in \Sect~9.

%############################################################

\section{Formation of shocks}

\subsection{Ballistic approximation and caustics}

Consider an outflow with internal supersonic motions. 
Such motions can be approximately described as ballistic:
each fluid element is moving with a constant velocity. It is well known that 
ballistic flows create caustics --- surfaces where density diverges
(in cosmology such surfaces are called Zeldovich pancakes).

The flow near the caustic is approximately plane-parallel 
(one-dimensional). It is convenient to view the flow in the rest frame of the caustic
and choose the $x$-axis normal to it, so that the flow converges toward $x=0$
along the $x$-axis with velocity $v(t,x)$.
Since $\beta=v/c$ may approach unity in relativistic flows, it is 
useful to introduce dimensionless momentum $p=\gamma\beta$
where $\gamma=(1-\beta^2)^{-2}$. Velocity is related to $p$ by
\beq
     v=\frac{c\,p}{(1+p^2)^{1/2}}.
\eeq

A simple example of a converging flow is provided by an initial $\arctan$ profile,
\beq
\label{eq:atan}
      p_0(x)=-\pmax\,\frac{2}{\pi}\arctan\frac{x}{\D}.
\eeq
The ratio $\pmax/\D$ describes the initial steepness of the wave, and 
$\pmax$ describes its amplitude. The wave is non-relativistic if $\pmax\ll 1$. 
The characteristic timescale of the profile evolution is $\tsh=\D/c\pmax$.
On this timescale the ballistic wave steepens (Figure~1)
and the caustic forms at $x=0$, where $-\partial v/\partial x$ is maximum. 

%%%%%%%%%%% FIGURE %%%%%%%%%%%%%%%%%%
\begin{figure}[t]
\begin{tabular}{c}
\includegraphics[width=0.44\textwidth]{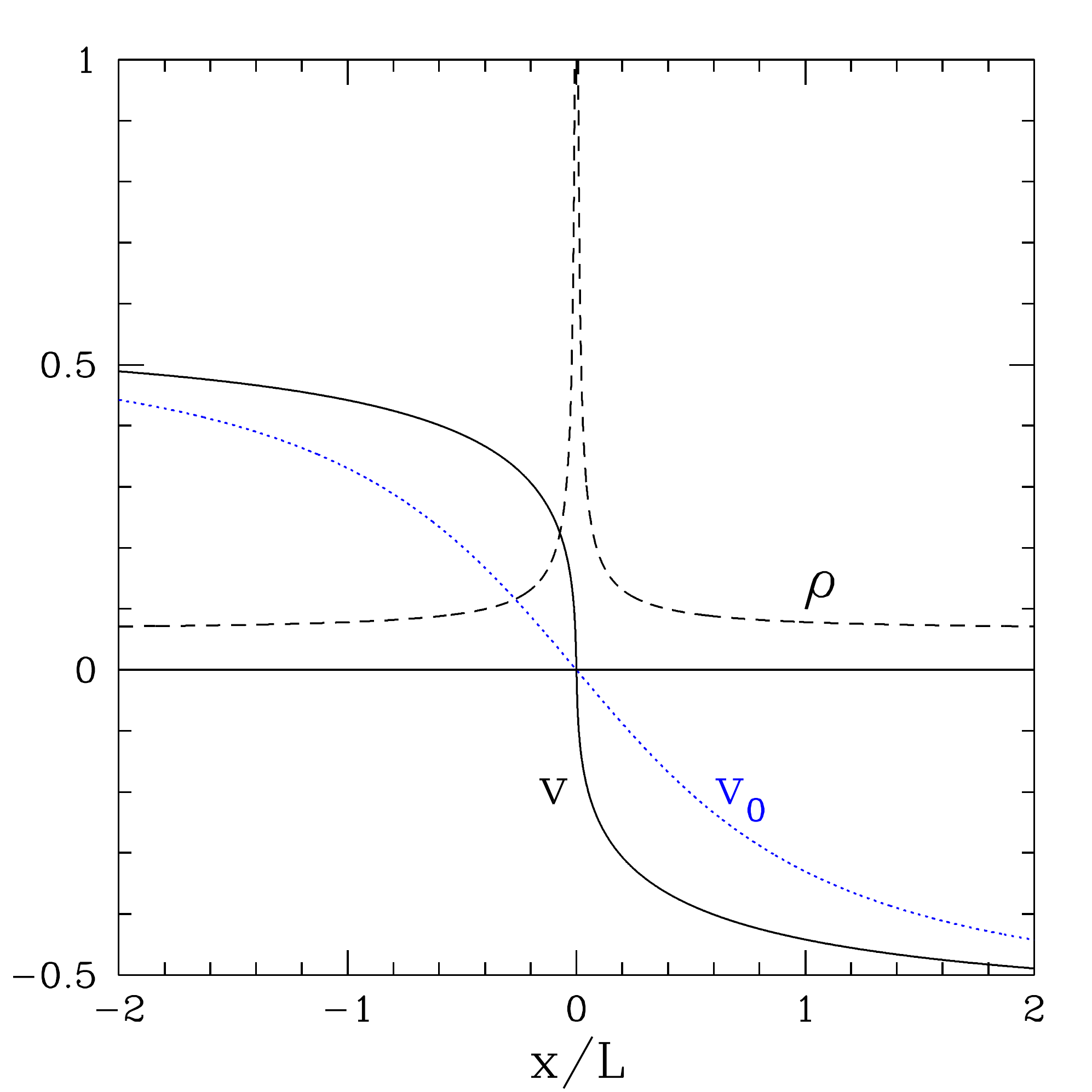}
\end{tabular}
\caption{Fluid velocity $v(x)$ (in units of $c$) and density $\rho(x)$
(arbitrary units) evolving in the ballistic compressive wave with the initial profile 
$v_0(x_0)$ given by \Eq~(\ref{eq:atan}), with $\pmax=0.7$.
The initial velocity profile is shown by the dotted curve, and the next 
snapshot is taken at the time of caustic formation $t_c\approx 2.24 L/c$. 
The dashed curve shows density $\rho$ at $t=t_c$; the initial density 
at $t=0$ was uniform.
 }
 \label{}
 \end{figure}
%%%%%%%%%%% FIGURE %%%%%%%%%%%%%%%%%%

The density of the ballistic flow diverges at the caustic. Its evolution is determined 
by the relation 
\beq
   x(x_0,t)=x_0+v_0t,
\eeq
where $x_0$ is the initial position of the fluid slab $dx_0$ at times $t_0\ll \tsh$,
and $v_0(x_0)=v_0(x)$ is the initial velocity profile. The slab $dx_0$ is contracting 
by the factor $(\partial x/\partial x_0)=1+(dv_0/dx_0)\,t$.
Therefore, the evolution of baryon density $\rho$ is described by
\beq
\label{eq:rho}
   \rho(x_0,t)=\frac{\rho_0(x_0)}{ 1 + v_0^\prime\,t },  \qquad v_0^\prime=\frac{dv_0}{dx_0},
\eeq 
where $\rho_0(x_0)=\rho_0(x)$ is the density at $t_0\ll \tsh$.
The compression rate is highest at $x=0$ and here density diverges at time
\beq
\label{eq:tsh}
  \tsh=-\left(\frac{d v_0}{d x_0}\right)_{x_0=0}^{-1}=\frac{\pi\D}{2c\pmax}. 
\eeq
At this moment, $v(x)$ becomes discontinuous at $x=0$.

\subsection{Pressure build-up in the converging flow}

True caustics form in flows with zero pressure.
A small initial pressure $P_0\neq 0$ qualitatively changes the picture:
it can be strongly amplified in the converging flow near $x=0$
and the generated pressure gradient stops the flow.

The deceleration of the converging flow around $x=0$ accelerates the steepening 
of the velocity profile on each side of the caustic  (Figure~2). As a result, at some 
time $\ts$ and locations $\pm \xs$  two shocks form and
continue to propagate away from $x=0$. 
The type of the nascent shock depends on the physical conditions in 
the region $x\sim \xs$. Below we discuss the pressure build up in the 
converging flow, then estimate the location of shock formation $\xs$ 
and the corresponding maximum compression.

%%%%%%%%%%% FIGURE %%%%%%%%%%%%%%%%%%
\begin{figure}[t]
\begin{tabular}{c}
\includegraphics[width=0.44\textwidth]{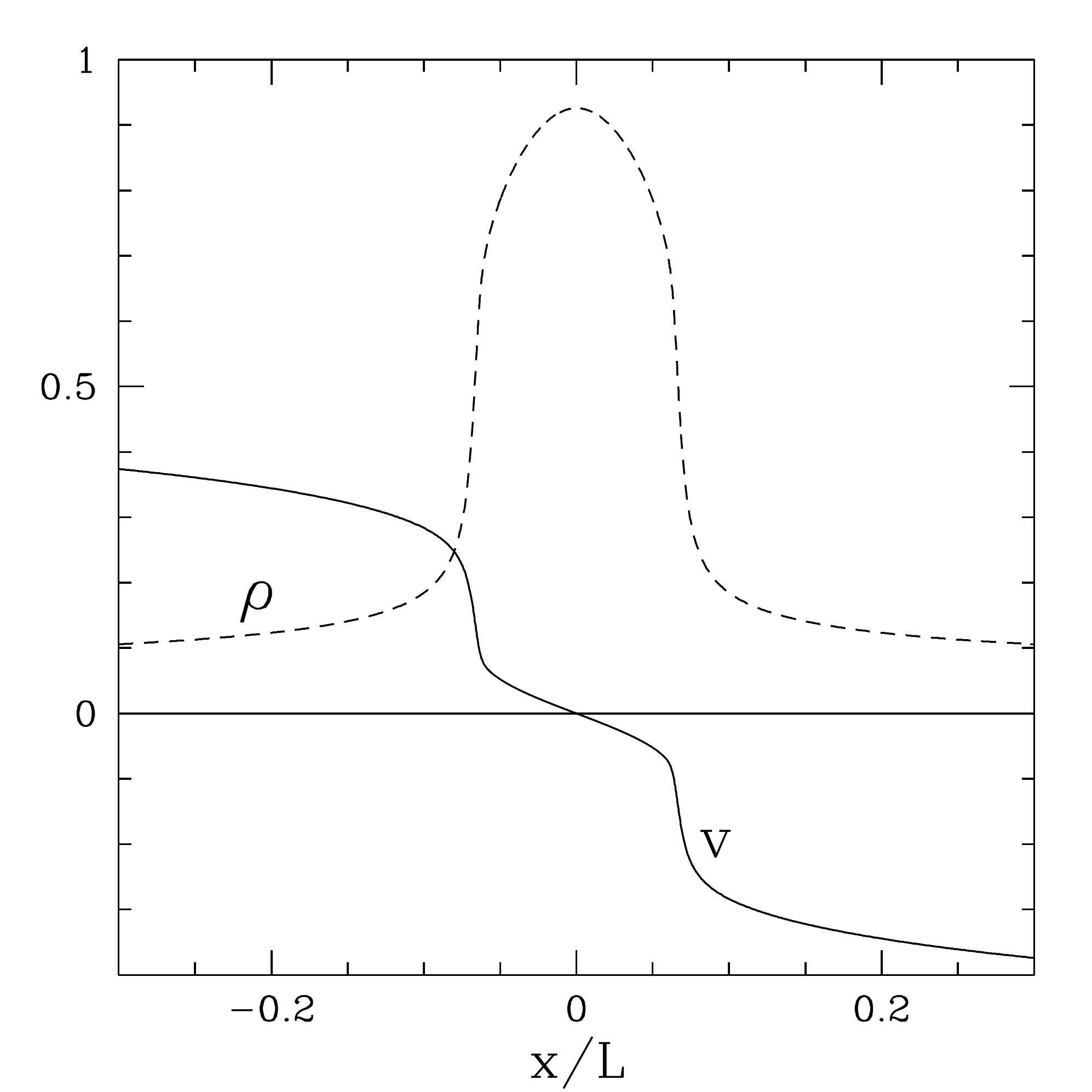} 
\end{tabular}
\caption{Snapshot of the evolved compressive wave with the same initial conditions
as in Figure~1 but with a finite initial pressure.
The flow has the initial sound speed $c_0=0.1c$ and adiabatic index $\alpha=4/3$. 
The snapshot is taken at $t\approx 2.8 L/c$, shortly before 
the formation of a pair of shocks. 
 }
 \label{}
 \end{figure}
%%%%%%%%%%% FIGURE %%%%%%%%%%%%%%%%%%

One source of pressure is the thermal motions of plasma particles.
It grows in the converging flow, however its contribution to the total pressure 
is limited by fast radiative cooling, which converts plasma heat to radiation. 
In a local thermodynamic equilibrium, radiation strongly dominates the heat 
capacity of GRB jets, because the photon density $n_\gamma$ greatly exceeds
the plasma density. At small radii, where the $e^\pm$ 
population is in annihilation equilibrium with Planck radiation, one finds
$n_\pm/n_\gamma\approx (kT/m_ec^2)^{-3/2}\exp(-2m_ec^2/kT)$ (Svensson 1984);
the $e^\pm$ abundance is decreasing exponentially in the expanding and adiabatically
cooling jet. At larger radii, the particle density (ions, $e^-$, or $e^+$) does not exceed 
$\sim 10^{-3}n_\gamma$ and, in a local thermodynamic equilibrium, this implies a plasma 
pressure $\Ppl\ll \Prad$. Here we examine compressive waves in a medium that is 
initially not too far from thermal equilibrium\footnote{Shocks 
    create strong deviations from thermodynamic equilibrium, and these deviations become
    long-lived in the region of moderate optical depth, around and above the photosphere. 
    New shocks in this region will develop in the plasma with hot (thermally decoupled) 
    ions, preheated by previous shocks. 
    }
and thus has $\Ppl\ll \Prad$. Then the two main sources of pressure that can be amplified 
in the wave are radiation and the transverse magnetic field.

Magnetic fields are expected to carry a significant fraction $\epsB$ of the jet energy.
Comparison of theoretical GRB spectra with observations suggests 
$\epsB\sim 0.01-0.1$ (Vurm \& Beloborodov 2016).
The jet plasma is an excellent conductor, so the magnetic field is frozen 
in it and advected by the flow.
In an internal compressive wave, the frozen transverse field is compressed together 
with the plasma:
$B\propto \rho$ or $\B\propto \rhof$, where $\B=B/\gamma$ is the magnetic field 
measured in the fluid frame, $\rhof=\rho/\gamma$ is the proper density of the fluid,
and $\gamma$ is its Lorentz factor. The magnetic pressure grows in the converging 
flow as\footnote{$P_B$ and $\rhof$ are measured in the same (fluid) frame.
    Pressure and internal energy density are always measured in the 
    fluid frame and we omit tilde to simplify notation.}
\beq 
\label{eq:PB}
  P_B=\frac{\B^2}{8\pi}\propto\rhof^2.
\eeq 

The growth of radiation pressure $\Prad$ depends on its ability 
to diffuse out if the compressed region, which depends on the optical depth. 
If the flow is sufficiently opaque to photons, the radiation will be trapped and 
\beq
\label{eq:Prad}
   \Prad=\frac{\Urad}{3}\propto \rhof^{4/3}   \qquad {\rm (trapped~radiation).}
\eeq
In the opposite limit, when the compressed region is transparent to photons, 
there is no significant amplification of $\Prad$.

\Eqs~(\ref{eq:PB}) and (\ref{eq:Prad}) both have a polytropic form $P\propto\rhof^{\alpha}$,
with $\alpha=2$ or $4/3$. A similar relation could also be used for compressive waves
in a medium that is far from thermal equilibrium with radiation and filled with hot, thermally 
decoupled, ions (protons). The proton pressure in the compressive wave follows the 
relation $P_p\propto \rhof^{\adind}$ with $\adind\approx 5/3$ as long as
the proton temperature is non-relativistic, $kT_p\simlt m_pc^2$.

\subsection{Shock formation in non-relativistic polytropic gas}

Let us first consider a non-relativistic gas, $P\ll\rhof c^2$.
Suppose that initially the gas has uniform pressure $P_0$ and density $\rhof_0$, and 
is set in motion with velocity $v_0(x)$ that corresponds to momentum profile
$p_0(x)$ given e.g. by \Eq~(\ref{eq:atan}).
We assume that the peak of velocity profile $v_{\max}=c\pmax(1+\pmax^2)^{-1/2}$ 
is much greater than the sound speed $c_0=(\adind P_0/\rhof_0)^{1/2}$. 
In the ballistic approximation, the profile would develop a caustic at $x=0$ at time $t_c$.
We wish to know how the finite pressure changes the flow dynamics, in particular what 
is the maximum pressure achieved in the compressed region before a shock forms, 
and where the shock formation occurs.

Even if the compressive wave is relativistic, $p_{\max}\simgt 1$, the condition $c_0\ll c$
implies that the shocks form not far from $x=0$ where $v_0(x)$ is non-relativistic.
Therefore, the shock formation can be examined using Newtonian hydrodynamics
around $x=0$, so we will use $\gamma\approx 1$, $\rhof\approx\rho$, and $\B\approx B$.

The evolution of the gas is convenient to view on the $x$-$t$ plane
(Figure~3). Each streamline is described by $x(x_0,t)$ where $x_0$ is the 
Lagrangian coordinate --- the position at $t=0$. 
Initially, a small fraction of gas is in the subsonic region $|v_0|<c_0$ near $x=0$.
The streamlines that start outside this region
are initially supersonic and eventually become subsonic. 

%%%%%%%%%%% FIGURE %%%%%%%%%%%%%%%%%%
\begin{figure}[t]
\begin{tabular}{c}
\includegraphics[width=0.45\textwidth]{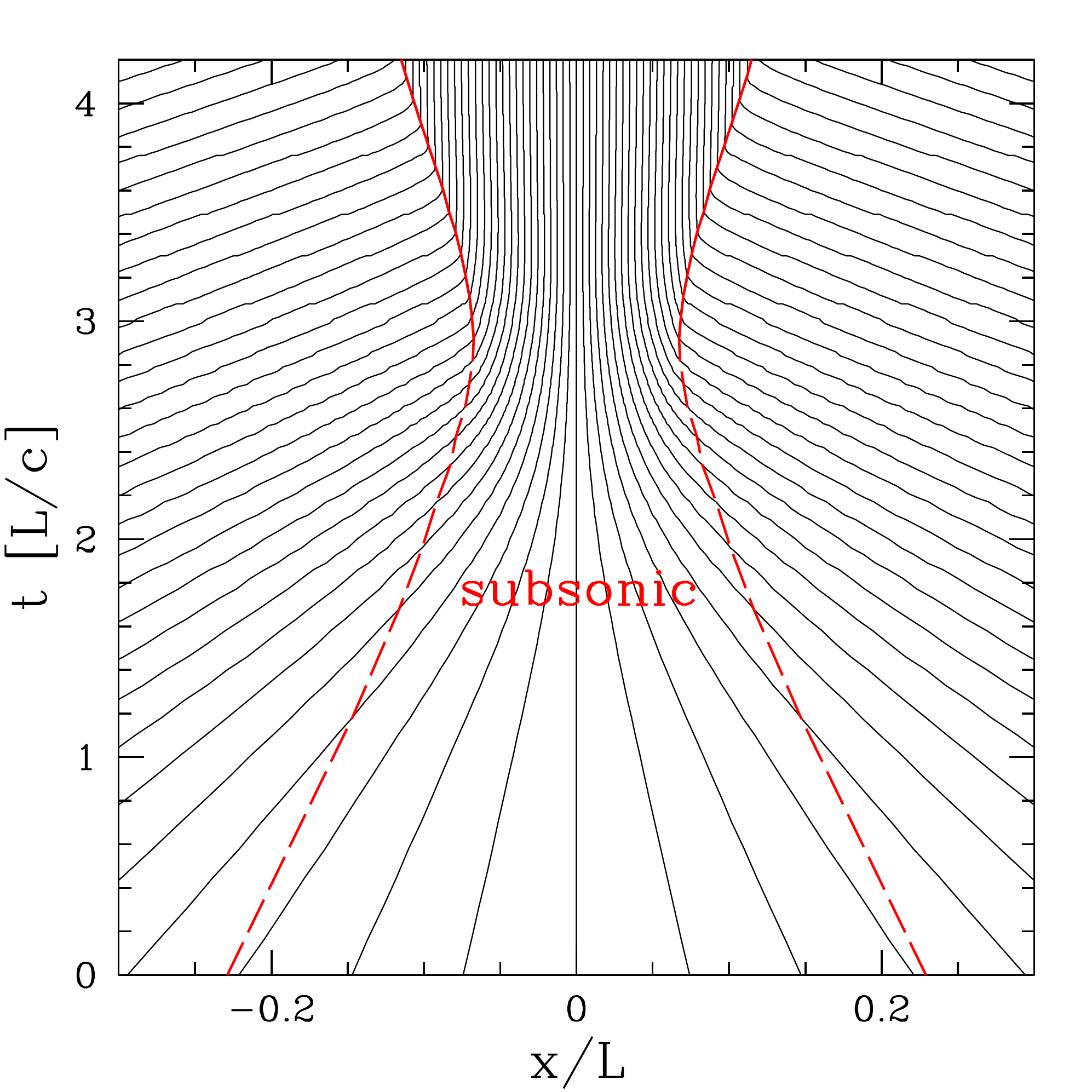} 
\end{tabular}
\caption{The streamlines on the spacetime diagram. The gas has the initial velocity profile 
$v_0(x_0)$ given by \Eq~(\ref{eq:atan}) with $\pmax=0.7$, the initial sound speed
$c_0=0.1c$, and the adiabatic index $\alpha=4/3$. The boundary of the subsonic region 
$v<c_s$ is shown by the red curves. The red curve is dashed where the deceleration to 
the subsonic speed occurs smoothly and solid where the deceleration occurs through a shock. The shock forms at $\ts\approx 2.9 L/c$ and $\xs\approx 0.07L$.
}
 \label{}
 \end{figure}
%%%%%%%%%%% FIGURE %%%%%%%%%%%%%%%%%%

There is a critical Lagrangian coordinate $\xx$.
Streamlines that start at $|x_0|<\xx$ will become subsonic without a shock:
the compressed gas is gradually decelerated as its specific kinetic energy $v_0^2/2$ 
gets transformed into enthalpy $(U+P)/\rho=c_s^2/(\alpha-1)$, 
where $c_s^2=c_0^2(\rho/\rho_0)^{\adind-1}$ is the local speed of sound. 
This ``compressive deceleration'' to a subsonic speed occurs when the 
compression factor $\rho/\rho_0$ satisfies
\beq
\label{eq:dec1}
    \frac{c_0^2}{\alpha-1}\left(\frac{\rho}{\rho_0}\right)^{\adind-1}\approx \frac{v_0^2}{2}.
\eeq
Approximating the streamline before this moment as ballistic, one can estimate
$\rho_0/\rho=1+\vp(x_0) t$.
Therefore, the deceleration time $\td(x_0)$ at which the streamline with Lagrangian 
coordinate $x_0$ becomes subsonic may be estimated from the condition,
\beq
\label{eq:td}
  (1+\vp\, \td)^{\adind -1} \approx \frac{2c_0^2}{(\alpha-1)v_0^2}.
\eeq
The corresponding location on the streamline is 
\beq
\label{eq:xd}
   \xd\approx x_0+v_0\, \td.
\eeq 
The smooth compressive deceleration is only possible 
for streamlines with sufficiently small $|x_0|<\xx$. For large $|x_0|$ one finds $\td>|x_0/v_0|$, 
and the compressive deceleration becomes impossible --- the ballistic flow does not 
have a chance to compress enough before it hits the existing subsonic region near $x=0$. 
Then the deceleration occurs through a shock. 

The critical Lagrangian coordinate $\xx$ at which the shock forms is given by 
(see Appendix~A),
\beq
\label{eq:xx1}
   \frac{\xx}{L}\approx 3 \left(\frac{c_0}{cp_{\max}}\right)^{1/\adind}.
\eeq
The compression of gas with the Lagrangian coordinate $\xx$ 
is determined by \Eq~(\ref{eq:dec1}) with $v_0$ evaluated at $\xx$. 
In the limit of $\xx\ll L$, the compression along this streamline is given by
\beq
 \frac{\rho_{\star}}{\rho_0}=\frac{\xx}{\xs}\sim \left(\frac{c\pmax}{c_0}\right)^{2/\adind}.
\eeq
It determines the maximum pressure developed in the flow before the shock is launched,
\beq
\label{eq:Ps}
  P_{\star}=P_0\left(\frac{\rho_\star}{\rho_0}\right)^\alpha
    \sim P_0\left(\frac{c\pmax}{c_0}\right)^2\sim \pmax^2\rho_0 c^2.
\eeq
The maximum pressure is comparable to the peak kinetic energy density of the wave,
even though $P_\star$ only develops in a small region near the caustic  $x\approx 0$
where the flow momentum is much smaller than $\pmax$.
This is because $P_\star$ is controlled by the {\it curvature} of the velocity profile 
(described by $v_0^{\prime\prime\prime}(0)$, see Appendix~A), which depends on 
$\pmax$.

At $\ts$ and $\xs$ the sound speed of the ballistic flow is not much below its
bulk speed $v\sim v_0(\xx)$, so the nascent shock is not strong. Then the shock 
propagates through the ballistic gas with increasing Lagrangian coordinate $|x_0|$ 
where the upstream velocity $v_0(x_0)$ is higher, and the shock 
compression ratio quickly approaches the strong-shock limit $(\adind+1)/(\adind-1)$.

\subsection{Shock formation in relativistic polytropic gas}

Shock formation in relativistic gas $P\gg\rhof c^2$ may be examined in a similar way.
This regime occurs in relativistic explosions at small radii where radiation dominates 
the gas inertia. Then $c_0^2=c^2/3$ (Landau \& Lifshitz 1959),
and one must consider relativistic compressive waves with $v_{\max}>c_0$.

The flow is initially subsonic in the zone where $|p_0|<2^{-1/2}$. 
Outside this zone the flow is approximately ballistic and its density is growing with 
time as $(1+\vp t)^{-1}$, where $\vp=c (dp_0/dx_0)\gamma_0^{-3}$.
The compressive deceleration of the relativistic gas is quite efficient:
a large fraction of the bulk kinetic energy is converted to enthalpy when the gas is 
compressed by only a factor of $\sim 2$. 

However, even such a moderate compression is difficult to achieve in the relativistic 
ballistic flow, because the gas with $\gamma_0\gg 1$ ($v_0\approx c$) has a small 
$\vp$ and hence it is compressed slowly. The maximum time allowed for ballistic 
compression is $x_0/c$ and the corresponding maximum compression factor is 
\beq
  (1+\vp |x_0|/c)^{-1}\approx 1 \qquad {\rm if} \quad |p_0|\gg 1.
\eeq
Gas with a relativistic $p_0$ ballistically hits the subsonic region
before it has a chance for compressive deceleration. Thus, the shock must form at 
Lagrangian coordinate $\xx$ such that $|p_0(\xx)|\sim 1$,
not far from the boundary of the initial subsonic zone $|p_0|=2^{-1/2}$. The time and 
location of shock formation are $\ts\sim\xx/c$ and $\xs\sim\xx/2$. The shock forms
with a mildly relativistic amplitude; it becomes ultra-relativistic when it propagates 
into the gas converging with $|p_0|\gg 1$.

One can also consider shock formation in a magnetically dominated gas
$P_B\gg \rhof c^2$ and $P_B\gg\Prad$. Then $c_0\approx c$ and it is convenient to define 
$\gamma_{c0}=(1-c_0^2/c^2)^{-1/2}$. Shocks form in compressive waves 
with $v_{\max}>c_0$, which corresponds to Lorentz factor 
$\gamma_{\max}>\gamma_{c0}$.

In the limit of strong magnetization, the sound speed becomes equal to $c$ and 
shocks do not form. In this case, the dynamic equations read 
$\partial_\mu T^{\mu\nu}=0$ with the stress-energy tensor components
\beq
   T^{tt}=\frac{B^2+E^2}{8\pi}, \quad T^{tx}=\frac{EB}{4\pi}, \quad T^{xx}=\frac{B^2+E^2}{8\pi},
\eeq
(the magnetic field $\bB$ is assumed to lie in the $y$-$z$ plane perpendicular to the fluid 
velocity).
The neglect of the plasma contribution to $T^{\mu\nu}$ defines so-called force-free 
electrodynamics, where plasma only serves to conduct electric currents demanded 
by $\nabla\times\bB$ and supplies no inertia. 
The plasma velocity $v=\beta c$ is related to the electric field by 
$\bE+\bv\times\bB/c=0$ and $\beta=E/B$. Adding and subtracting the 
energy and momentum conservation laws, 
\beq
   \frac{\partial T^{tt}}{\partial t} +c\,\frac{\partial T^{tx}}{\partial x} =0, \qquad
   \frac{\partial T^{tx}}{\partial t} +c\,\frac{\partial T^{xx}}{\partial x} =0,
\eeq
one obtains
\beq
    \frac{\partial u_{\pm}}{\partial t}\pm c\frac{\partial u_\pm}{\partial x}=0, \qquad
    u_\pm=\left(1\pm\beta\right) B.
\eeq 
The initial profiles of $u_\pm(0,x)=f_\pm(x)$ determine $u_\pm(t,x)=f_\pm(x\mp ct)$.
This gives explicit solutions for $B=(u_++u_-)/2$ and $\beta=(u_+-u_-)/(u_++u_-)$,
demonstrating their smooth behaviour, with no caustics or shocks.

%################################################################

\subsection{Radiation diffusion and formation of RMS}

Radiation diffusion is an essential ingredient of an RMS, since it is the mechanism
of shock propagation. However, too fast diffusion would let radiation escape,
inhibiting the RMS formation.  A shock wave is usually assumed to be
radiation-mediated if two conditions are satisfied:
\medskip

\noindent
(A) The jump conditions give in the downstream $P\approx\Prad$, so that a large 
fraction of energy generated by the shock is carried by radiation (Zeldovich \& Raizer 1966).
\medskip

\noindent
(B) The medium has optical depth $\tau>c/v_0$, so that the shock generates radiation 
faster than it could diffuse out of the system of size $L$. For instance, in a supenova 
explosion one could take $L$ as the radius of the expanding ejecta (e.g. Tolstov et al. 
2013). 
\medskip

\noindent
In fact, these conditions do not guarantee that the shock is mediated by radiation. 
The velocity profile connecting the upstream and downstream may 
contain a ``subshock'' --- a sharp jump mediated by the plasma
on a scale much smaller than the photon free path to scattering. 
In non-relativistic shocks ($v_0\ll c$) satisfying conditions (A) and (B) the velocity 
profile is smooth, with no subshock (Zeldovich \& Raizer 1966). 
However, in the relativistic case, $v_0\approx c$, a weak subshock was reported 
(Budnik et al. 2010). In addition, in the above condition (B) one should be careful with
what is meant by the ``size of the system.'' 

Consider a compressive wave of a mildly relativistic amplitude $v_{\max}\sim c$ and 
length $\sim L$. A characteristic optical depth may be defined as
\beq
   \tau_L=L\rho_0\kappa,
\eeq
where $\kappa$ is the opacity of the gas.
Suppose the unperturbed gas has a non-relativistic sound speed $c_0\ll  c$. 
\Sect~2.3 described how at time $\ts\sim L/c$ two shocks form 
near the caustic, at the Lagrangian coordinate $\xx/L\sim (c_0/c)^{1/\adind}$.
Thus, the region of shock formation has the optical depth 
\beq
\label{eq:taus}
   \taus\sim \tau_L\left(\frac{c_0}{c}\right)^{1/\adind}.
\eeq
For a radiation-dominated flow, $P\approx \Prad$ and $\adind=4/3$.
In this case, however, \Eq~(\ref{eq:taus}) can only be used if radiation is trapped,
i.e. unable to diffuse out of the region of shock formation on the timescale 
$\xs/v_0^{\star}$. This requires 
\beq
\label{eq:trap}
    \taus\gg \frac{c}{v_0^{\star}},
\eeq 
where $v_0^\star\approx c\,\taus/\tau_L$ is the flow velocity upstream of the nascent shock. The trapping condition is satisfied if 
\beq
\label{eq:trap1}
   \frac{c_0}{c}\gg \tau_L^{-2/3}.
\eeq
For the flow with the upstream pressure $P_0$ dominated by radiation, one can use the 
relation 
\beq
\label{eq:eps}
     \frac{c_s}{c}\approx \left(\frac{w}{3}\right)^{1/2},
     \qquad w\equiv \frac{\Urad+\Prad}{\rhof c^2}.
\eeq
Then condition~(\ref{eq:trap1}) may also be written as $w\gg 3\tau_L^{-4/3}$.
If this condition is satisfied, a propagating jump in radiation pressure will develop at $\ts$,
and the nascent shock will be mediated by photons, i.e. an RMS will be launched.

The RMS velocity profile is shaped by the competition between advection of radiation 
through the shock and its diffusion in the opposite, upstream direction.
Therefore, the optical depth of the velocity jump $\taush$ is regulated to 
\beq
\label{eq:dtau}
   \taush\sim \frac{c}{v_0}.
\eeq
The RMS propagation involves continual amplification of radiation 
advected through the shock --- the result of photon 
scattering in the region of a steep velocity gradient.
As the hot downstream photons diffuse back into the upstream, they experience 
``bulk Comptonization'' --- they are boosted in energy by the factor of 
$\sim\gamma_0^2=(1-v_0^2/c^2)^{-1}$.
As a result, the energy of radiation advected through the shock is amplified,
as required by the jump conditions for a propagating shock.

Launching an RMS at $\ts$ requires an initial build-up of radiation density 
$\sim \rho {v_0^\star}^2/2$ near the shock front, which is only possible if 
the trapping condition~(\ref{eq:trap1}) is satisfied. Otherwise,
 radiation leaks out of the compressed region to large distances $x$.
This may be viewed as a violation of RMS condition (B), as the effective 
``size of the system'' during the shock formation is comparable to $\xs\ll L$.
Then the radiation pressure gradient is too weak to control the velocity 
profile of the flow. Radiation is unable to resist the steepening of the velocity profile, 
and the width of the velocity jump is quickly reduced to the ion Larmor radius,
forming a collisionless shock mediated by the collective electromagnetic field. 
It may later evolve into an RMS, when the postshock region has accumulated 
a sufficient optical depth, if there is enough time for that in the expanding outflow,
i.e. if the shock forms sufficiently deep below the photosphere.

\subsection{Critical magnetization}

When both magnetic field and radiation contribute to pressure, 
there are two contributions to the sound speed, $c_s^2=c_{\rm rad}^2+c_{B}^2$.
It is convenient to define the dimensionless enthalpy of the flow,
\beq
    w=\frac{\Prad+\Urad}{\rhof c^2}=\frac{4\Prad}{\rhof c^2}.
\eeq
A similar quantity for the magnetic field $\B$ in the fluid frame is 
\beq
   \sigma=\frac{P_B+U_B}{\rhof c^2}=\frac{\B^2}{4\pi\rhof c^2}.
\eeq
Suppose the optical depth is large so that the radiation trapping condition is satisfied.
The type of the nascent shock is determined by whether the magnetic field or radiation 
dominates the pressure in the compressed region near the caustic, 
$P_\star=\Prad^\star+P_B^\star$. An RMS forms if 
$P_\star$ is dominated by radiation; otherwise, a collisionless shock is launched.
Since $P_\star$ must be the same in either case (see \Eq~\ref{eq:Ps}), it is sufficient
to compare the compressions needed to reach $P_\star$ with only magnetic or only 
radiation pressure: $(P_\star/\Prad)^{3/4}$ and $(P_\star/P_B)^{1/2}$. 
This comparison gives an approximate condition for launching a collisionless 
shock in a cold ($w_0\ll 1$) and opaque medium,
\beq
\label{eq:cl}
   \sigma_0 >\frac{w_0^{3/2}}{4\pmax}.
\eeq
Note that this condition only applies to the nascent shock, at the point of maximum 
ballistic compression in the converging wave near the caustic. As the shock becomes
stronger and continues 
to propagate into the ballistic flow where the upstream is less compressed (and hence
less magnetized), its type may change. 

An established steady shock structure is determined by the parameters of its upstream.
The first step in the analysis of a steady propagating shock is the solution for its 
jump conditions.

%##############################################

\section{Shock jump conditions}

Jump conditions for relativistic magnetized shocks were studied by
de Hoffmann \& Teller (1950). Pulsar wind nebulas and GRBs revived interest to 
relativistic shocks.
The upstream medium is usually assumed to be cold in the sense that its enthalpy 
is much smaller than the rest-mass energy of the plasma. This condition 
is not, however, satisfied in the inner regions of GRB jets. Below we write down the 
general jump conditions for shocks propagating in a hot magnetized plasma filled with 
radiation, and show their solutions. 

Consider a shock wave propagating in a sufficiently extended, optically thick medium.
Far upstream and far downstream of the shock the plasma and radiation can be described
as an ideal gas with isotropic pressure. 
The thermal energy density $U$ and pressure are related by
\beq
  U=\frac{P}{\alpha-1},
\eeq
where $\alpha=4/3$, as long as $U$ is dominated by radiation.
When formulating the jump conditions we will keep $\alpha$ general, and
specialize to $\alpha=4/3$ in numerical solutions. 

In the rest frame of the upstream (pre-shock) fluid, an observer will see the
downstream (post-shock) fluid approaching with velocity $\bv_0$. 
The shock front is perpendicular to $\bv_0$ and approaching with a higher 
velocity $\bv_1$ parallel to $\bv_0$. The plasma carries a frozen magnetic field 
$\B$ (measured in the fluid rest frame). 
We only consider magnetic fields perpendicular to the fluid velocity;
a parallel magnetic field is anyway unchanged by the shock and hence does not affect
the jump conditions.

\subsection{Stress-energy tensor and sound speed}

The stress-energy tensor of a hot magnetized flow with four-velocity 
$u^\mu=(\gamma c,\gamma{\mathbf v})$ is given by
\begin{eqnarray}
\nonumber
    T^{\mu\nu} &=& (\rhof c^2+U+P)\,\frac{u^\mu u^\nu}{c^2}+g^{\mu\nu}P  \\
                      &+&  \frac{1}{4\pi}\left(F^{\sigma\mu}F_{\sigma}^\nu
       -\frac{g^{\mu\nu}}{4}F^{\sigma\delta}F_{\sigma\delta}\right),
\end{eqnarray}
where $\rhof$ is the proper rest-mass density of the baryons, 
$g_{\mu\nu}$ is the metric tensor of Minkowski spacetime, and $F^{\mu\nu}$ is the 
electromagnetic tensor. 
Its electric and magnetic components in the lab frame, $\bE$ and $\bB$, are related by
${\mathbf E}+{\mathbf v}\times{\mathbf B}/c=0$, as the plasma is a nearly ideal conductor. 
Using $\bv\perp\bB$ and $B=\gamma\B$, the stress-energy tensor may be reduced to the 
ideal fluid form, 
\beq
\label{eq:T}
   T^{\mu\nu}=\Heff\,\frac{u^\mu u^\nu}{c^2}+g^{\mu\nu}\Peff,
\eeq
with the effective relativistic enthalpy and pressure 
\beq
   \Heff=\rhof c^2+U+P+\frac{\B^2}{4\pi}, \qquad \Peff=P+\frac{\B^2}{8\pi}.
\eeq

Before considering shocks, it is useful to examine sound waves in a 
uniform background that has $u_0^\mu=(c,0,0,0)$, $\Heff^0=const$, and $\Peff^0=const$. 
Let us consider longitudinal (compressive) waves propagating along the $x$-axis. 
In the linear order, perturbations are described by the four-velocity $u^\mu=(c,v,0,0)$,
four-acceleration $u^\mu\nabla_\mu u^\nu=(0,\partial_t v,0,0)$, and 
compression $\nabla_\mu u^\mu=\partial_x v$. The linearized equations of motions 
$\nabla_\mu T^{\mu\nu}=0$ give (for $\nu=t$ and $\nu=x$) 
\begin{eqnarray}
   \frac{\partial \Heff}{\partial t} + \Heff^0\,\frac{\partial v}{\partial x}-\frac{\partial \Peff}{\partial t}=0, \\
   \Heff^0\,\frac{\partial v}{\partial t}+c^2\,\frac{\partial \Peff}{\partial x}=0.
\end{eqnarray}
These two equations can be reduced to the wave equation for $v$,
\beq
   \frac{\partial^2 v}{\partial t^2} - c_s^2\,\frac{\partial^2 v}{\partial x^2}=0, 
\eeq
where the wave speed $c_s$ is defined by
\beq
\label{eq:s}
   \qquad  c^2\,d\Peff=c_s^2\,d\left(\Heff-\Peff\right).
\eeq
As the two main parameters of the fluid, it is convenient to use the dimensionless 
contributions of enthalpy and magnetic fields to the fluid inertia, 
\beq
     w\equiv\frac{U+P}{\rhof c^2}, \qquad \a\equiv \frac{\B^2}{4\pi\rhof c^2}.
\eeq 
Then the wave speed defined in \Eq~(\ref{eq:s}) may be expressed as
\beq
\label{eq:cs}
   \frac{c_s^2}{c^2}=\frac{(\alpha-1)w+\sigma}{1+w+\sigma}.
\eeq
This general expression reduces to familiar cases in four limits:\\ 
(1) $\sigma\ll w\ll 1$: $c_s^2=\alpha P/\rho$ 
(non-relativistic sound waves),\\
(2) $w\ll\sigma\ll 1$: $c_s^2=B^2/4\pi \rho$ 
(non-relativistic fast MHD modes in a cold plasma),\\
(3) $w\gg\sigma,1$: $c_s^2=(\adind-1)c^2=c^2/3$ (sound waves in a relativistic gas), and\\
(4) $\sigma\gg w,1$: $c_s=c$ (force-free limit of the MHD modes).

Internal supersonic motions $v_0>c_s$ generate shocks, as discussed in detail in \Sect~2. 
In addition, shocks can form through nonlinear steepening of sound waves excited by 
a subsonic perturbation, $v_0<c_s$ (Zeldovich \& Raizer 1966).
The steepening occurs because $c_s$ is slightly increased in the region compressed 
by the wave, so the crest of the wave (maximum $v>0$ and maximum $\rho$) 
travels faster than the trough (minimum $v<0$ and minimum $\rho$).
Using \Eq~(\ref{eq:cs}) one can verify that $dc_s/d\rhof>0$, i.e. compression indeed
increases the local sound speed.
The shock formed through sound-wave steepening propagates super-sonically but has 
a subsonic velocity jump, i.e. it separates regions with a relative velocity $v_0<c_s$.
Such weak shocks are found among the solutions shown below, along with strong shocks 
formed by supersonic motions $v_0>c_s$.

Formation of shocks through steepening of sound waves is inefficient 
in the relativistic regimes (3) and (4), as in this case $dc_s/d\rhof\rightarrow 0$
($c_s=c/\sqrt{3}$ or $c_s=c$ is constant in both cases).
In the force-free limit ($w\ll \sigma\gg 1$), shock formation does not occur at all
(\Sect~2.4). In a radiation-dominated medium ($\sigma\ll w\gg 1$), shocks can be 
launched by a supersonic motion $v_0>c/\sqrt{3}$.

\subsection{Jump conditions}

Jump conditions express the continuity of fluxes of energy, momentum, and
baryon number in the rest frame of the shock front. The fluxes of energy and momentum 
are given by the stress-energy tensor $T^{\mu\nu}$ in \Eq~(\ref{eq:T}).
The baryon flux is described by the four-vector
\beq
    F^\mu=\rhof u^\mu.
\eeq
The fluxes along the shock normal (the $x$-axis) are given by
\begin{eqnarray}
\label{eq:Txx}
\label{eq:Ttx}
   T^{tx} & = & \gamma p\,\rhof c^2\left(1+w+ \sigma\right), \\
   T^{xx} & = & p^2\rhof c^2\left(1+w+\sigma\right)+P+\frac{\B^2}{8\pi},\\
    F^x &=& p\, \rhof\, c,
\end{eqnarray}
where $p=\gamma\beta$, and $P$ can be expressed in terms of $w$: 
$P=(1-\adind^{-1})w\rhof c^2$.
Equating the fluxes upstream (index ``u'') and downstream (index ``d'') one obtains the relations,
\begin{eqnarray}
\label{eq:j1}
  \frac{T^{tx}}{cF^x}&=&\gd\left(1+\wdo+\sigd\right)
                                    =\gu\left(1+\wu+\sigu\right), \\
\nonumber
 \frac{T^{xx}}{cF^x}&=& \pd\left(1+\wdo+\sigd\right)
                                       +\left(1-\frac{1}{\adind}\right) \frac{\wdo}{\pd}+\frac{\sigd}{2\pd} \\
                           &=& \pu\left(1+\wu+\sigu\right)+\left(1-\frac{1}{\adind}\right)\frac{\wu}{\pu}+\frac{\sigu}{2\pu}.
\label{eq:j2}
\end{eqnarray}
Given the upstream parameters $\pu$, $\wu$, $\sigu$, and
taking into account that $\pd\sigd=\pu\sigu$ (implied by the flux freezing condition $\B\propto\rhof$),
one can solve \Eqs~(\ref{eq:j1}) and (\ref{eq:j2}) for the two unknowns $\pd$ and $\wdo$.

Typically, the upstream velocity relative to the downstream, $v_0$, is a given in the shock 
problem. Therefore, we chose $p_0=\gamma_0 v_0/c$ as an independent parameter 
instead of $\pu$.
The upstream momentum in the shock frame, $\pu$, is related to the upstream 
momentum measured in the downstream frame, $p_0$, by the Lorentz transformation
between the two frames,
\beq
\label{eq:p1}
   \pu=\gd(p_0+\bd\gamma_0).
\eeq
For a given $p_0$, a trial $\pd$ determines $\pu(\pd)$, and the solution of 
\Eqs~(\ref{eq:j1}) and (\ref{eq:j2}) 
(which is obtained numerically) yields $\pd$ and $\pu$ together with $\wdo$.

The solutions are shown for $p_0=1$ and 10 in Figures~4 and 5,
assuming $\adind=4/3$ (pressure is dominated by radiation).
Figure~4 shows the compression ratio $\xi=\rhofd/\rhofu$ (which also 
determines $\sigd=\xi \sigu$) and Figure~5 shows the ratio $\wdo/\sigd$. 
The latter determines the dissipation efficiency of the shock: if $\wdo/\sigd>1$ 
then a large fraction of the upstream energy goes to heat rather than ends up stored in 
the compressed magnetic field. The ratio $\wdo/\sigd$ is also interesting for another 
reason: it is related to the dissipation mechanism in the shock front, as discussed below.

%%%%%%%%%%% FIGURE %%%%%%%%%%%%%%%%%%
\begin{figure}[t]
\begin{tabular}{c}
\includegraphics[width=0.45\textwidth]{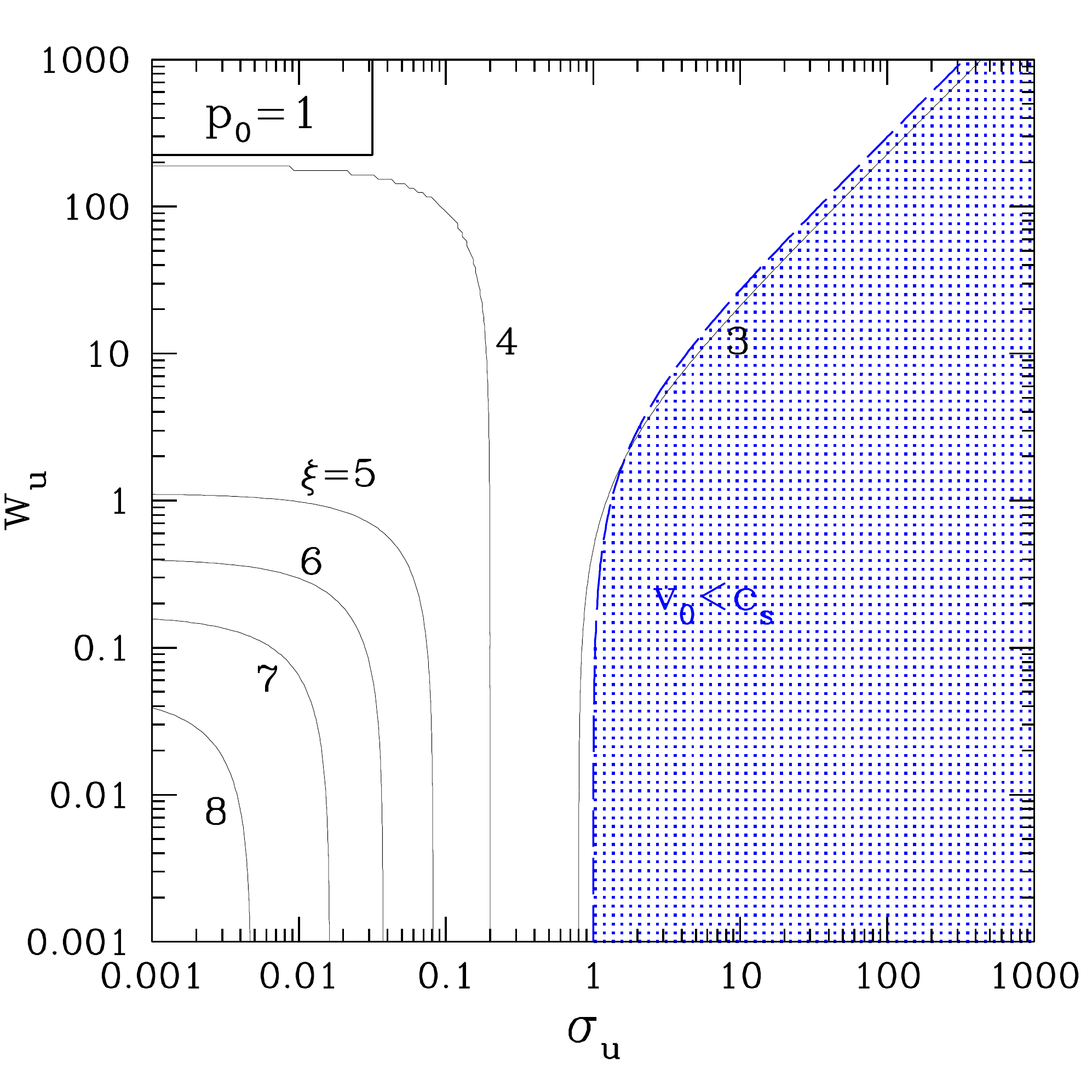} \\
\includegraphics[width=0.45\textwidth]{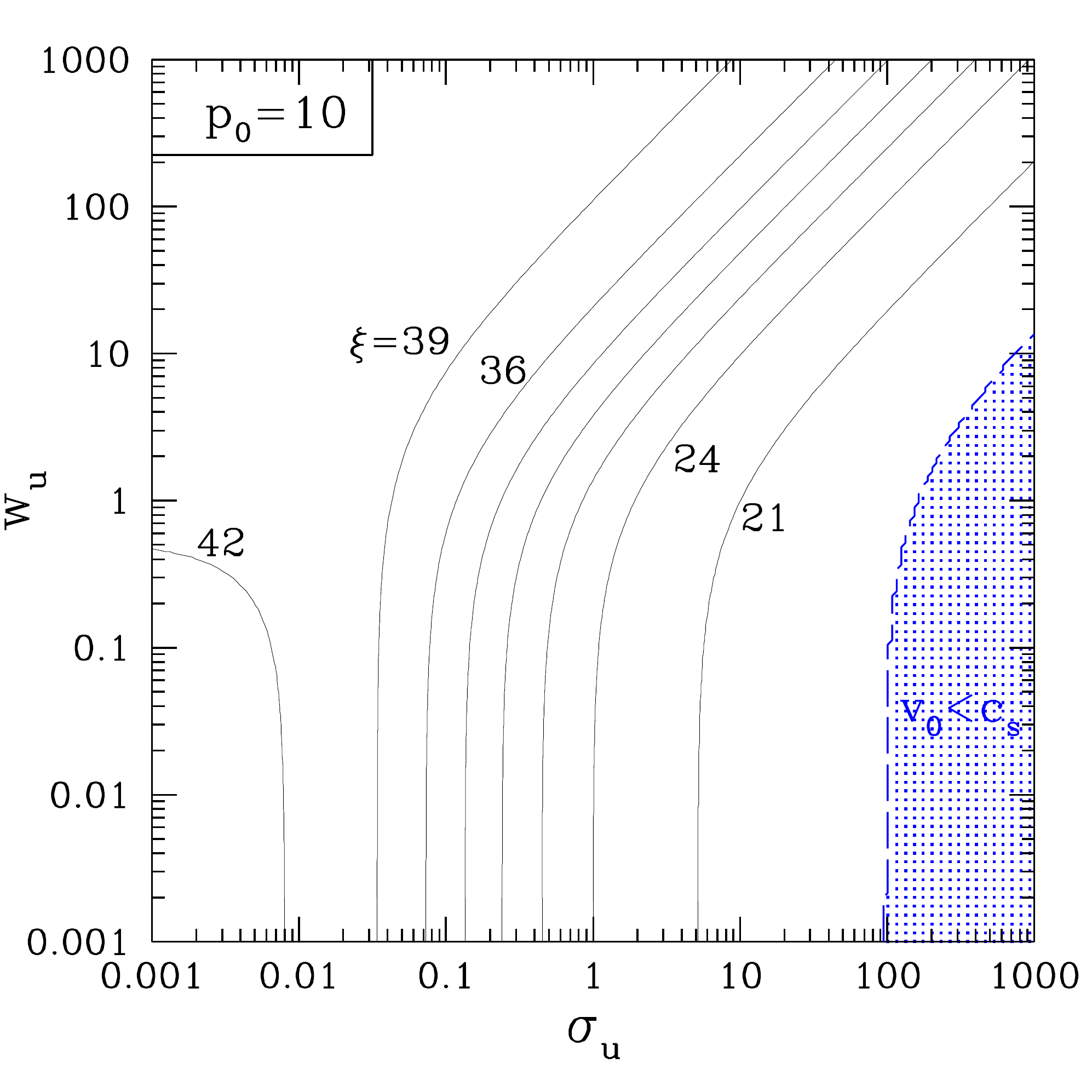} 
\end{tabular}
\caption{Compression ratio $\xi=\rhofd/\rhofu$ in a relativistic shock with $p_0=1$ 
(upper panel) and $p_0=10$ (lower level). The compression $\xi$ depends on the 
upstream enthalpy $\wu$ and magnetization $\sigu$. Solid curves show contours 
of the function $\xi(\sigu,\wu)$.
The region where the shock is weak, $v_0<c_s$, is dotted in blue and bounded by 
the dashed curve.
}
 \label{fig:jc_xi}
 \end{figure}
%%%%%%%%%%% FIGURE %%%%%%%%%%%%%%%%%%
%%%%%%%%%%% FIGURE %%%%%%%%%%%%%%%%%%
\begin{figure}[t]
\begin{tabular}{c}
\includegraphics[width=0.45\textwidth]{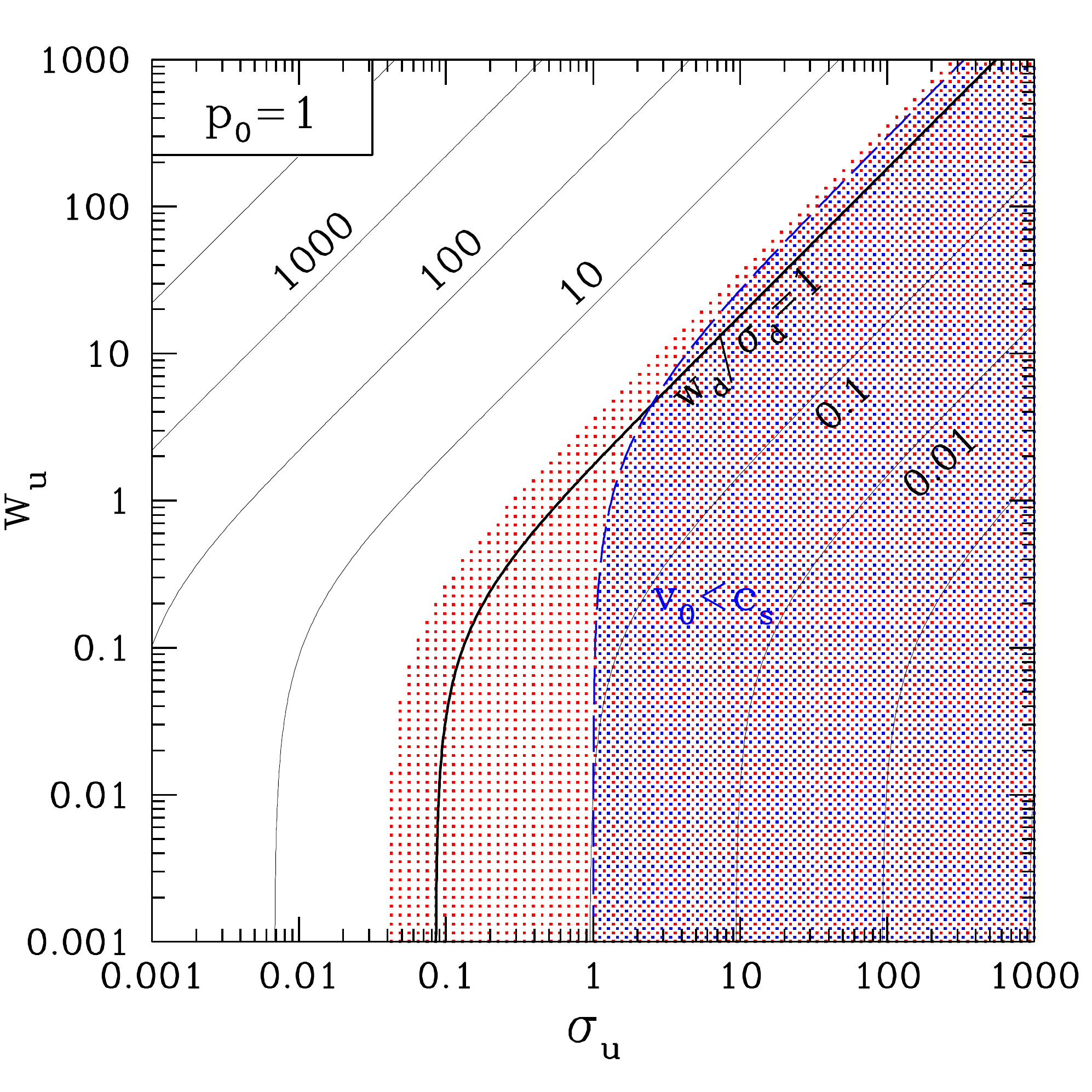} \\
\includegraphics[width=0.45\textwidth]{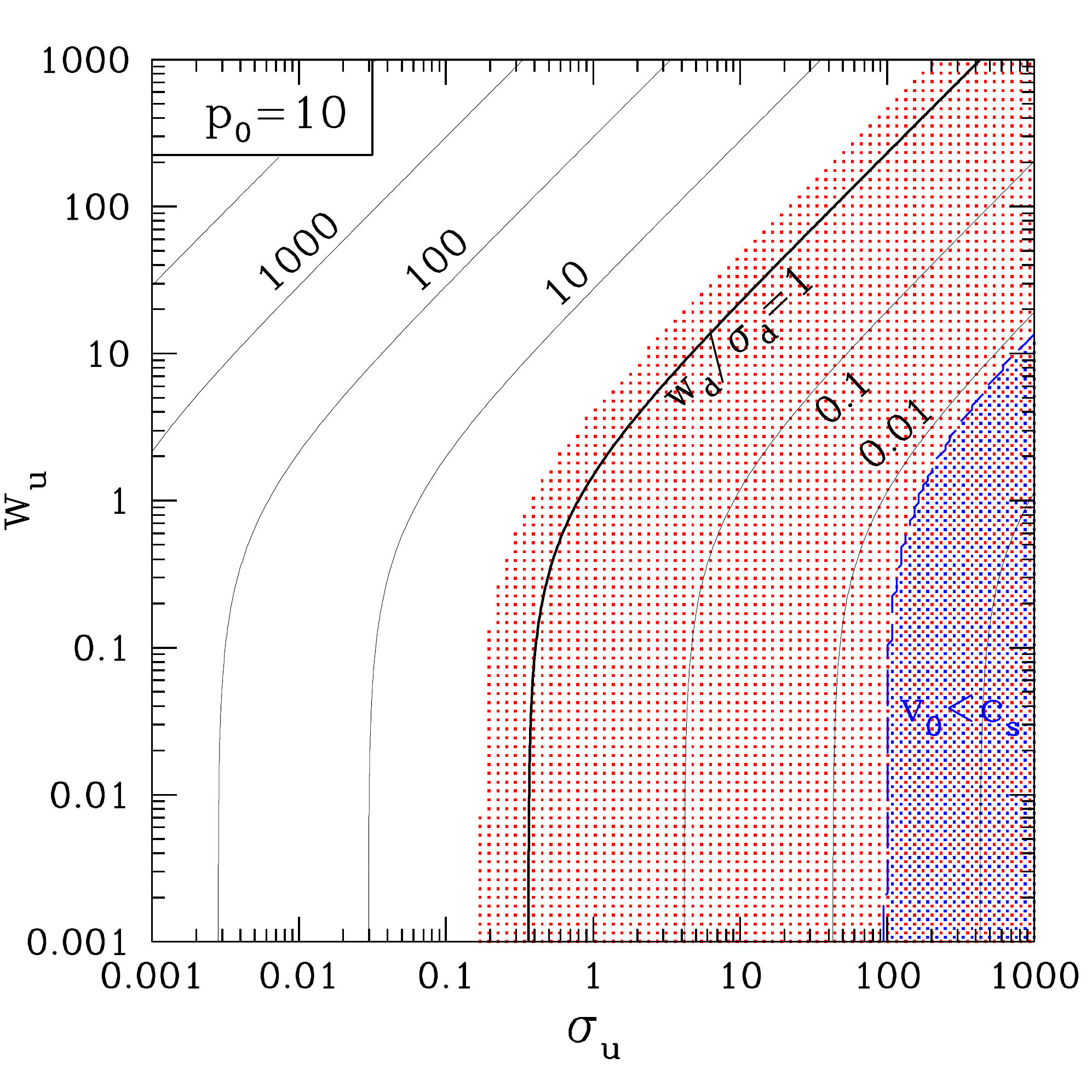} 
\end{tabular}
\caption{Ratio of the enthalpy and magnetization in the downstream, $\wdo/\sigd$, 
for shocks with $p_0=1$ (upper panel) and $p_0=10$ (lower panel). This ratio depends 
on the upstream enthalpy $\wu$ and magnetization $\sigu$; the dependence is 
shown using contours on the $\sigu$-$\wu$ plane. The blue region is the same 
as in Figure~4. Red dots highlight the region 
where $\wdo<2\sigd$. In this region, the shock is expected to have a strong collisionless 
jump where the velocity changes on a scale comparable to the gyroradius of the 
plasma particles.
}
 \label{fig:jc_ratio}
 \end{figure}
%%%%%%%%%%% FIGURE %%%%%%%%%%%%%%%%%%

\subsection{Collisionless shocks}

The jump conditions do not describe the structure or dissipation mechanism of the 
shock front. However, they allow one to evaluate the region in the parameter space 
where dissipation must be mainly collisionless. 

We expect the shock to be mainly mediated by collective electromagnetic fields when the 
downstream enthalpy $\wdo$ is dominated by the compressed magnetic field $\sigd$. 
Then radiation cannot control the shock structure, as its pressure
is below the ram pressure of the shock. In particular, in the limit of $\wdo\ll\sigd$ 
the downstream can be approximated as a cold magnetized medium with negligible heat, 
so radiation has no effect on the upstream deceleration and the shock velocity profile;
the profile inevitably steepens so that the entire velocity jump occurs in a thin layer 
on the collisionless plasma scale (gyroradius). 

In contrast, in shocks with significant ratio $\wdo/\sigd$, 
the diffusion of the postshock radiation into the upstream region creates a precursor
that changes the upstream velocity and reduces the amplitude of the collisionless jump.
The resulting structure may be described as a collisionless shock with a radiation 
precursor or, equivalently, an RMS with a collisionless subshock. 
In the limit of $\wdo\gg\sigd$, the subshock becomes weak or non-existent. 
\Sect~4 below demonstrates this fact for shocks in 
relativistic gas, $w\gg 1$), and \Sect~5 will show the shock structure for a moderate 
$w=0.1$ with and without a significant magnetic field.

The transition between the two dissipation regimes --- mainly mediated by collective 
electromagnetic fields and mainly mediated by radiation --- occurs at $\wdo/\sigd\sim 2$
(the shock structure in this transition region will be calculated in \Sect~5).
The region where a strong collisionless jump is expected ($\wdo/\sigd\simlt 2$) 
is highlighted in red in Figure~5. We also show (in blue) the region where the 
shock is weak ($v_0<c_s$) and could only form through steepening of sound waves.
As discussed in \Sect~3.1, such shocks do not easily form in a relativistic fluid
($w>1$ or $\a>1$) since steepening takes a long time, typically longer that the 
expansion time of the outflow. The region between the two curves 
$\wdo=2\sigd$ and $v_0=c_s$ is where strong collisionless shocks occur.
When $\wdo\ll\sigd$ only a small fraction of the upstream kinetic energy 
is dissipated in the shock, and most of it ends up stored in the compressed 
magnetic field. Therefore, the strongest collisionless dissipation is expected
if $\wdo\sim\sigd$. 

For a medium with a given enthalpy $\wu$ one can define a characteristic magnetization 
$\sigu=\sigcr$ such that a shock propagating in the medium will have $\wdo/\sigd=2$
(the boundary of the red-dotted region in Figure~\ref{fig:jc_ratio}).
The magnetization $\sigcr$ depends on the shock strength $p_0$ 
(upstream momentum measured in the downstream frame). 
This dependence is shown in Figure~6.
In the ultra-relativistic limit $p_0\gg 1$, we find that $\sigcr$ does not depend on $p_0$
(both $\wdo$ and $\sigd$ scale as $p_0^2$, so their ratio does not depend on $p_0$). 
For non-relativistic shocks $p_0\ll 1$ with a cold upstream ($\wu=0$), $\sigcr$ scales as 
$p_0^2$. This is because weakly magnetized shocks have downstream enthalpy 
$\wdo\propto p_0^2$ while $\sigd\propto\sigu$.

%%%%%%%%%%% FIGURE %%%%%%%%%%%%%%%%%%
\begin{figure}[t]
\begin{tabular}{c}
\includegraphics[width=0.45\textwidth]{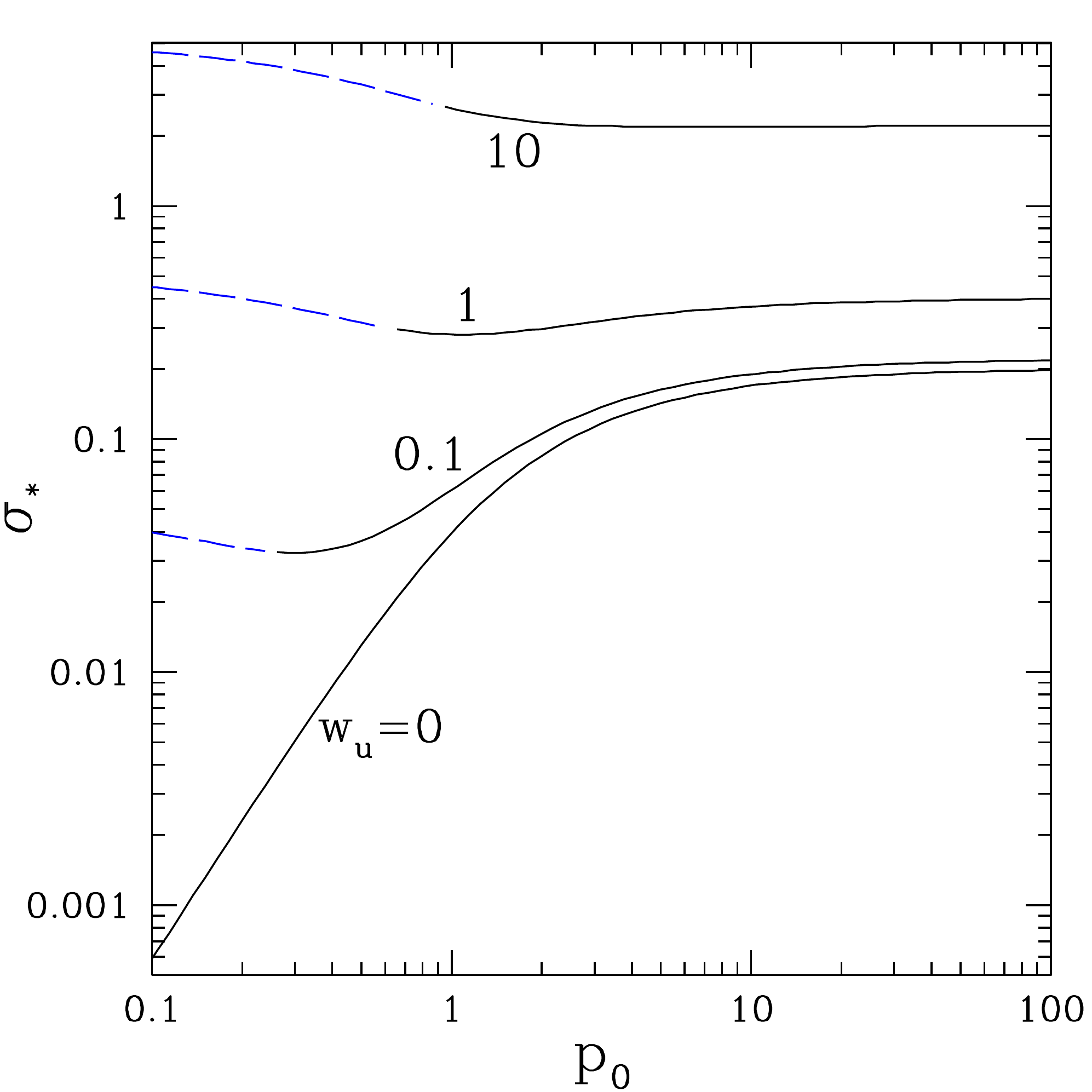}
\end{tabular}
\caption{The characteristic upstream magnetization $\sigcr$ that gives 
$\wdo/\sigd=2$ in the downstream. 
A strong collisionless subshock should form in the RMS if $\sigu\simgt\sigcr$.
Each curve shows the dependence of $\sigcr$
on the shock strength $p_0$ for a fixed upstream enthalpy $\wu$ (indicated next
to the curve). The curves are dashed where the shock is weak, $v_0<c_s$.
}
 \label{}
 \end{figure}
%%%%%%%%%%% FIGURE %%%%%%%%%%%%%%%%%%

At large $\sigma>1$ and $\sigma>w$, the effective sound speed $c_s$ approaches 
the speed of light (see \Eq~\ref{eq:cs}).
Shocks easily form if the internal compressive motions are supersonic,
i.e. their Lorentz factors $\gamma_0$ exceed $\sigma^{1/2}$. 
The dissipation in magnetically dominated shocks 
occurs in a microscopically thin, collisionless shock front.
Dissipation is reduced in this regime, 
as a large fraction of shock energy goes into the compressed
magnetic field, however dissipation can still be significant.
For example, a shock with Lorentz factor $\gamma_0=4$ propagating in a 
medium with upstream enthalpy $w=1$ and $\sigma=10$ has the 
downstream enthalpy $\wdo\approx 6.8$ and $\Urad\approx 0.12\,U_B$.
In this example, the shock compression ratio is $\xi\approx 8.3$, and
adiabatic compression would imply the amplification of $w$ by only a factor of 
$\xi^{1/3}\approx 2$, well below 6.8. 
We note also that the downstream enthalpy $\wdo$ is mainly determined by the 
upstream enthalpy $\wu$ and the amplitude of the shock; it weakly depends on $\sigu$.

%##############################################

\section{Shocks in photon gas}
\label{sec:photon_gas}

In GRB explosion models, sub-photospheric shocks begin to form at early stages, 
when the jet rest mass is still dominated by radiation, before the jet accelerates to 
its asymptotic Lorentz factor. This section examines the structure of shock waves 
in this regime, neglecting magnetic fields.

In essence, we deal here with shocks in the gas of photons, as the plasma inertia is 
negligible. The plasma role is to provide opacity and thus to couple the photons into 
a single fluid, with a small mean free path. The plasma particles may be viewed as 
passive ``markers'' following the motion of the photon gas.

\subsection{Jump conditions}

Far upstream and far downstream of the shock, the radiation can be described as 
ideal fluid with isotropic pressure $P$ and the stress-energy tensor 
\beq
    T^{\mu\nu}=4Pu^\mu u^\nu+Pg^{\mu\nu},
\eeq
where we have used the equation of state $U=3P$. Jump conditions
express the continuity of $T^{tx}$ (flux of energy) and $T^{xx}$
(flux of momentum) in the rest frame of the shock,
\begin{eqnarray}
\label{eq:jc1}
   4\Pd\gd^2\bd &=& 4\Pu\gu^2\bu, \\
   4\Pd\gd^2\bd^2+\Pd &=& 4\Pu\gu^2\bu^2+\Pu,
\label{eq:jc2}
\end{eqnarray}
where subscript ``u'' stands for upstream and ``d'' for downstream; pressure is 
measured in the fluid frame, and velocity is measured in the shock frame.
Dividing \Eqs~(\ref{eq:jc1}) and (\ref{eq:jc2}), one finds that $\bu$ and $\bd$
satisfy the condition
\beq
   g(\bu)=g(\bd), \qquad  g(\beta)=\frac{1+3\beta^2}{\beta}.
\eeq 
Rewriting the definition of $g$ as 
\beq
   3\beta^2-g\beta+1=0,
\eeq
one can view $\bu$ and $\bd$ as the two roots of the quadratic equation, and hence
they are related by
\beq
\label{eq:jcbeta}
   \bu\bd=\frac{1}{3}.
\eeq
Since $\bu>\bd$, one concludes that $\bu>3^{-1/2}$.
This condition merely states that the shock moves supersonically
relative to the upstream (recall that the sound speed is $c_0=3^{-1/2}c$).
Using the relation~(\ref{eq:jcbeta}) and \Eq~(\ref{eq:jc1}) or (\ref{eq:jc2}), 
one finds the pressure jump across the shock
\beq
\label{eq:jcP}
   \frac{\Pd}{\Pu}=3\gu^2\left(\bu^2-\frac{1}{9}\right).
\eeq
The shock compresses the volume measured in the fluid frame by the factor
\beq
\label{eq:xi}
   \xi=\frac{\gu\bu}{\gd\bd}=\gu\bu\left(9\bu^2-1\right)^{1/2},
\eeq
which also gives the relation
\beq
\label{eq:Pratio}
   \frac{\Pd}{\Pu}=\frac{\xi^2}{3\bu^2}.
\eeq
For ultra-relativistic shocks, $\bu\rightarrow 1$, the jump conditions simplify to 
$\bd=1/3$ and $\Pd/\Pu=(8/3)\gu^2=\xi^2/3$.

\subsection{Evolution equation for the photon gas}

Formation of shocks in the gas of photons can be simulated numerically.
It is convenient to think of this problem
as a radiative transfer problem for the bolometric intensity of radiation $I$.
Since the stress-energy tensor is dominated by radiation, the plasma 
is effectively massless and its velocity $\beta$ is controlled by the ``force-free'' condition:
$\beta$ equals the equilibrium value such that the radiation flux in the fluid frame 
vanishes (zero flux implies zero force applied by radiation to the plasma).
This condition leads to a well-defined radiative transfer 
problem (Beloborodov 1999). It has a simple solution for steady spherically 
symmetric relativistic outflows (Beloborodov 2011). Here we are interested in shock 
formation in variable outflows, so the problem is time-dependent. 

The shock is thin and locally flat (in the $y$-$z$ plane),
and we can study its formation in the plane-parallel geometry.
Then the bolometric intensity is described by the function $I(t,x,\mu)$ 
where $\mu=\cos\theta$ and $\theta$ is the photon angle with respect to the $x$-axis.

The stress-energy tensor of radiation is determined by the moments of the intensity,
\beq
    I_k(t,x)=\frac{1}{2}\int_{-1}^{1} I(t,x,\mu)\,\mu^k\,d\mu.
\eeq
In particular, $T^{tt}=4\pi I_0$, $T^{tx}=4\pi I_1$, and $T^{xx}=4\pi I_2$.
The force-free condition reads $\If_1=0$ in the fluid frame, and
the transformation of $T^{\mu\nu}$ from the lab frame to the fluid frame gives 
the quadratic equation for velocity,
\beq
    \If_1=\gamma^2\left[-\beta(I_0+I_2)+(1+\beta^2)I_1\right]=0,
\eeq
\beq
\label{eq:beta}
 \Rightarrow \qquad  \beta=\zeta-\left(\zeta^2-1\right)^{1/2}, 
     \qquad \zeta\equiv\frac{I_0+I_2}{I_1}.
\eeq

The evolution of intensity is described by the transfer equation,
\beq
\label{eq:transfer}
   \frac{1}{c}\,\frac{\partial I}{\partial t}=-\mu\,\frac{\partial I}{\partial x}
      +n\sT(1-\beta\mu)(S-I),
\eeq
where $n$ is the number density of electrons/positrons measured in the lab frame,
and $S$ is the source function. In the simplest case of isotropic scattering, $S$
is given by (Beloborodov 1999)
\beq
    S(\mu)=\frac{I_0-\beta I_1}{\gamma^4(1-\beta\mu)^4}.
\eeq

\subsection{Numerical solution}

\begin{figure}[t]
\begin{tabular}{c}
\includegraphics[width=0.4\textwidth]{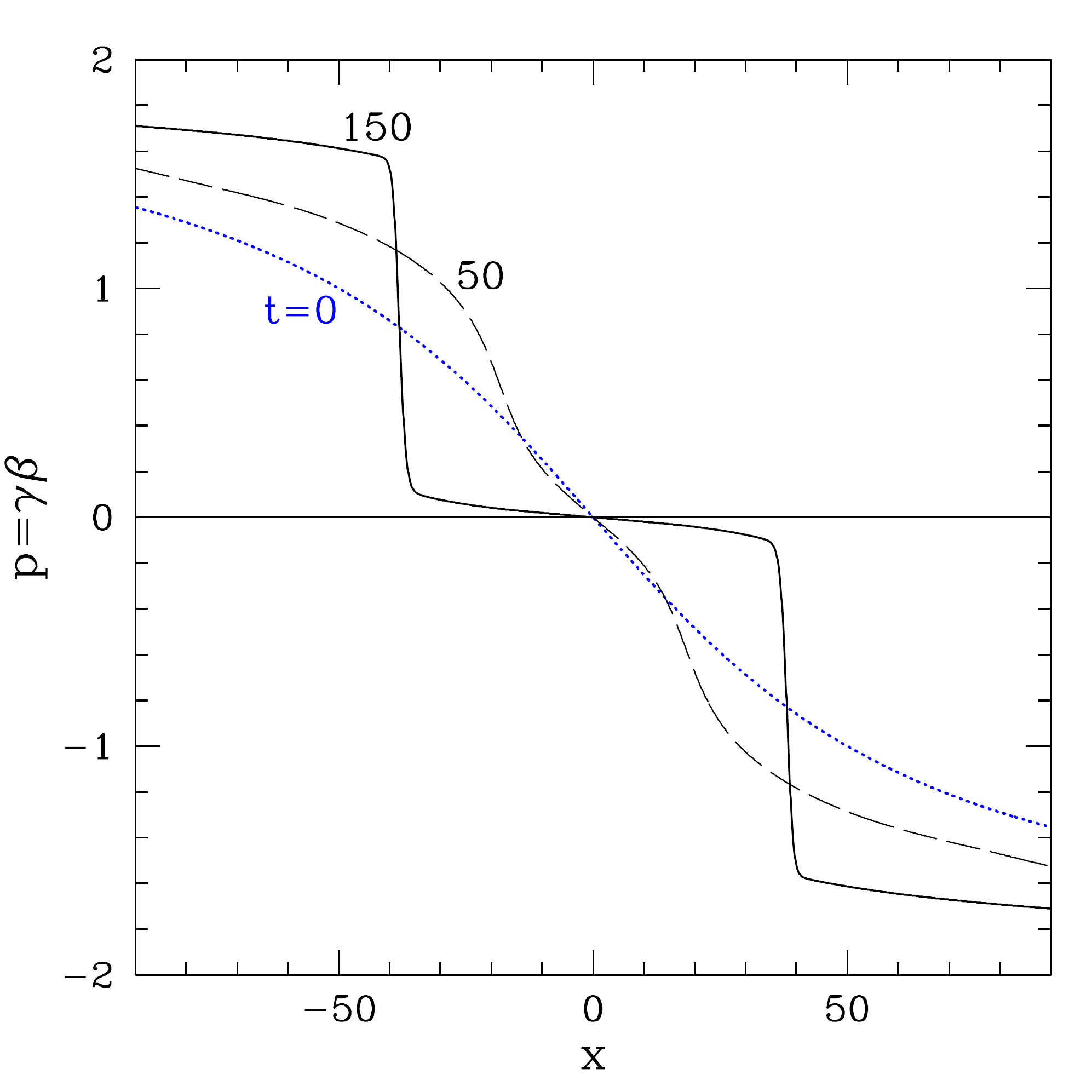} \\ 
\includegraphics[width=0.4\textwidth]{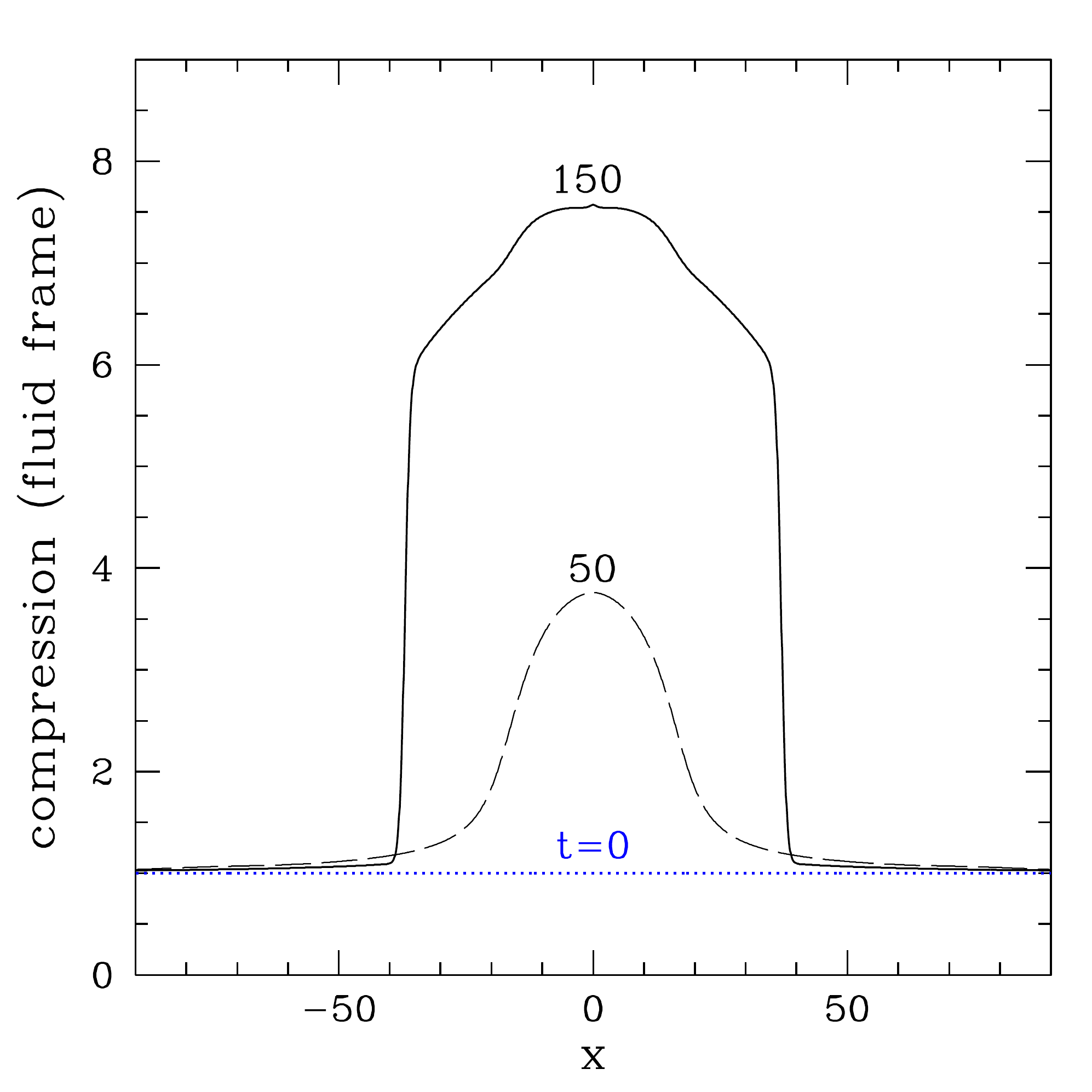} \\
\includegraphics[width=0.4\textwidth]{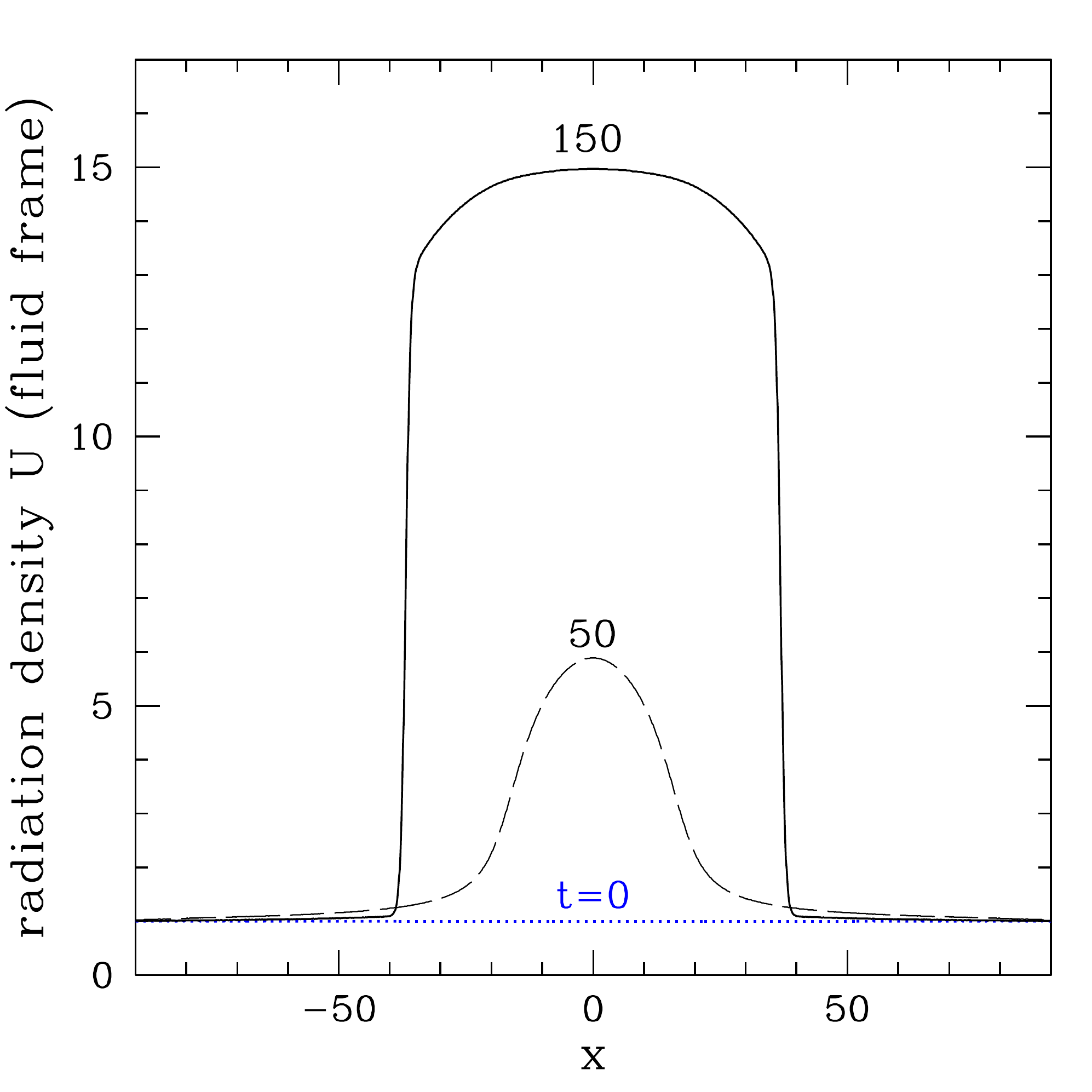}
\end{tabular}
\caption{Evolution of the fluid velocity, compression, and radiation 
energy density $U$ (measured in the fluid frame). 
Unit length in $x$ corresponds to a slab of unit Thomson optical depth in the initial state. 
The indicated times are measured in units where $c=1$.
}
 \label{fig:evol}
 \end{figure}

\Eq~(\ref{eq:transfer}) supplemented with the equilibrium velocity condition 
(\Eq~\ref{eq:beta}) can be solved numerically.
A sample solution is shown in Figure~\ref{fig:evol}. 
In this example, the initial state is given by \Eq~(\ref{eq:atan})
with $\pmax=2$ and $L=50$. The initial plasma density is uniform in the lab frame, 
$\rho(0,x)=\rho_0=const$,  
and the unit length in $x$ corresponds to a slab of unit Thomson optical depth.
The initial radiation density measured in the fluid frame is uniform in the lab frame,
$U(0,x)=U_0=const$. 

%%%%%%%%%%% FIGURE %%%%%%%%%%%%%%%%%%
%%%%%%%%%%% FIGURE %%%%%%%%%%%%%%%%%%

%%%%%%%%%%% FIGURE %%%%%%%%%%%%%%%%%%
\begin{figure}[h]
\begin{tabular}{c}
\includegraphics[width=0.44\textwidth]{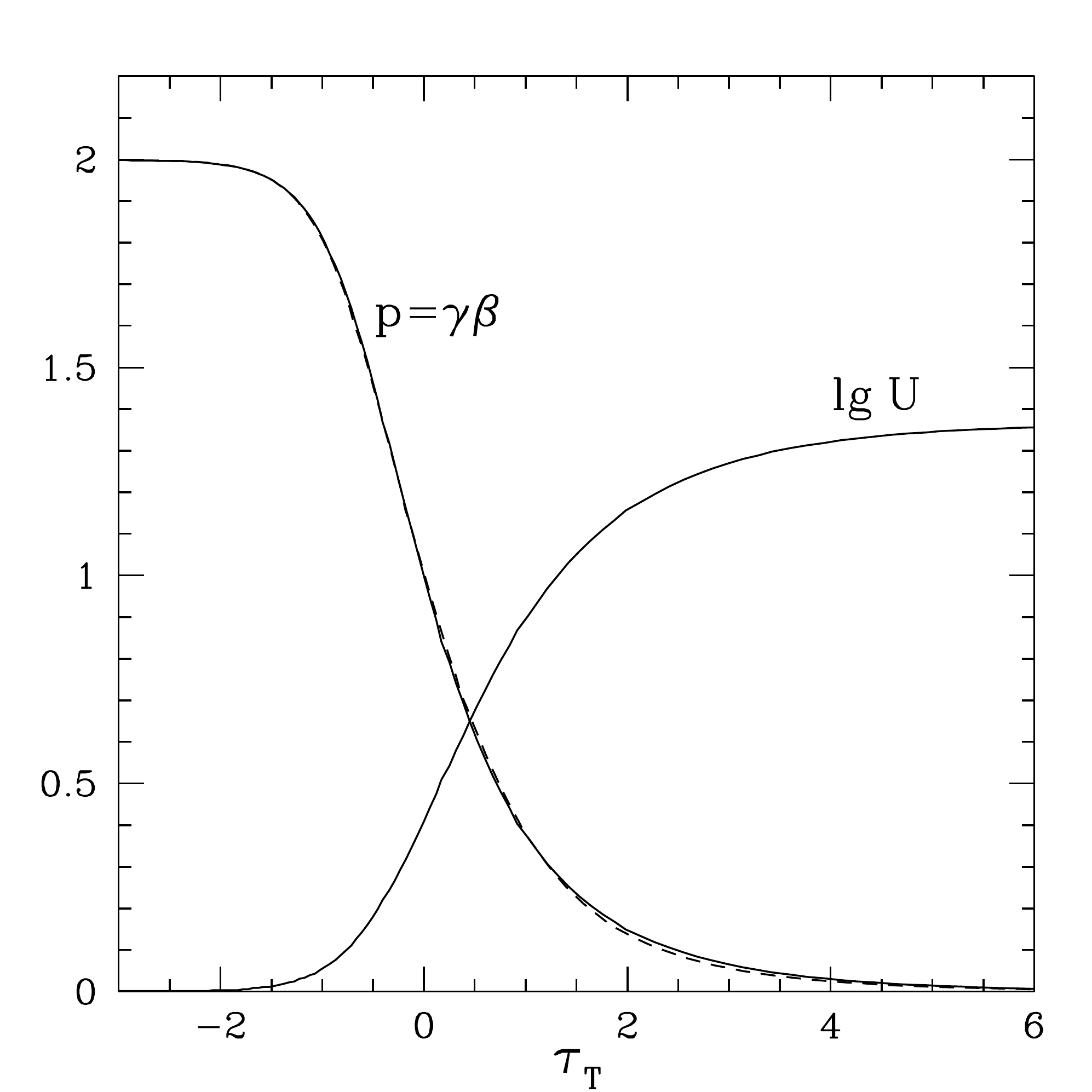}
\end{tabular}
\caption{Shock wave in ``photon gas'' ($\Urad\gg\rhof c^2,\, U_B$).
The shock structure is shown as a function of Thomson optical depth $\tauT$ 
in the frame where the downstream is at rest. The reference point $\tauT=0$ 
is chosen at half-maximum of the momentum profile $p=p_0/2$. 
Energy density $U=4\pi\If_0$ is measured in the fluid frame
(normalized to its pre-shock value).
The solid curves show the shock wave with amplitude $p_0=2$. 
The dashed curve shows $p(\tauT)/2$ for the shock with $p_0=4$.
}
 \label{fig:rms}
 \end{figure}
%%%%%%%%%%% FIGURE %%%%%%%%%%%%%%%%%%

One can see the compression of the converging supersonic flow
and the formation of a pair of shocks symmetric about $x=0$, as described in \Sect~2.
As the two shocks continue to propagate,
the downstream fluid comes nearly to rest in the lab frame.
The upstream velocity relative to the downstream, $\beta_0$, is related 
to the upstream and downstream velocities measured in the shock frame by 
\beq
   \beta_0=\frac{\bu-\bd}{1-\bu\bd}.
\eeq
Using $\bd\bu=1/3$ (\Sect~4.1), one finds
\beq
   \bu=\frac{1}{3}\left[\beta_0+(\beta_0^2+3)^{1/2}\right].
\eeq

Once the shock wave is established, its structure becomes independent of the details of 
the initial conditions. The shock has only one parameter: $\beta_0$ 
or $p_0=\gamma_0\beta_0$.
Figure~\ref{fig:rms} shows the obtained structure of a shock wave with $p_0=2$.
It is shown as a function of the optical depth measured in the $x$ direction.
Then the result is independent of the plasma density, so the obtained solution is unique.
Using $\beta_0=2^{-1/2}$ and $\bu=0.7903$ that correspond to $p_0=2$, 
one finds from \Eqs~(\ref{eq:xi}) and (\ref{eq:Pratio}) 
the ratio of downstream and upstream pressures $\Pd/\Pu=\Ud/\Uu=23.3$.
This asymptotic value of $U$ is observed in Figure~\ref{fig:rms}.

For comparison, Figure~\ref{fig:rms} also shows the momentum profile $p(x)$ 
for a shock with $p_0=4$, obtained from a similar time-dependent simulation. 
When re-scaled by the factor of $2$, the momentum profile is the same as for $p_0=2$.

%###############################################################

\section{Monte-Carlo simulations of shocks}

Time-dependent simulations may also be employed to study shocks in plasma 
with significant rest mass and magnetic fields. In contrast to the photon gas studied
in \Sect~4, now
the radiative transfer equation cannot be closed by the force-free condition $\If_1=0$. 
Instead, the fluid acceleration must be calculated together with the radiative transfer.
Another complication is the need to follow the evolution of the radiation 
spectrum, which develops a hard tail extending above $\sim m_ec^2$ inside the shock
front; then the scattering cross section is changed by the electron recoil.

Below we solve this 
problem using a direct numerical experiment that follows individual photons
and their interactions with the plasma,
so that the transfer of  momentum and energy is described on a microscopic level. 
In the initial state, the flow has a smooth velocity profile described by \Eq~(\ref{eq:atan})
and carries thermal radiation, which is isotropic in the fluid frame.
The flow is opaque and supersonic, which
leads to the formation of a pair of shocks propagating in the $\pm x$ directions, 
as described in \Sect~2. 

The flow has two interacting components:
\\
(1) Magnetized plasma. The plasma is assumed to carry a transverse magnetic field, 
which provides strong coupling between all charged particles, so their dynamics along 
the $x$ axis is well described as a single-fluid motion.\footnote{For any realistic
   magnetic field the Larmor radii of ions and electrons are microscopic, many orders 
   of magnitude smaller than the photon free path. Therefore, we assume strong 
   magnetic coupling of the plasma particles even in the weakly magnetized model 
   that is formally labeled as $\sigma=0$.
   }
In the numerical simulations we use a Lagrangian grid moving together with 
the fluid of charged particles: the fluid is discretized into 
$N\sim 10^4$ shells of equal rest mass $m$ and a small scattering optical depth.
Besides mass, each shell is characterized by the magnetic flux frozen in it,
internal thermal energy, and total pressure. The magnetic flux remains constant
while thermodynamic quantities may change as the 
shell contracts (or expands) and interacts with radiation. Thermal conductivity of 
the plasma in the $x$-direction is suppressed by the transverse 
magnetic field and neglected.
\\
(2) Radiation. Radiation is represented by $\sim 10^8$ photons which are followed 
individually. The photons migrate through the plasma shells and 
occasionally scatter off a thermal electron.
The scattering is followed using Monte-Carlo technique,
with the exact Klein-Nishina differential cross section and assuming that the 
thermal electrons are isotropic in the fluid frame.
The electrons are assumed to have a Maxwellian distribution with a
self-consistently calculated temperature.

A detailed description of the numerical method will be given in an upcoming paper.
A possible alternative to the Monte-Carlo method is the solution of the transfer equation 
for the radiation intensity, as in \Sect~4 but now including fluid inertia. A similar approach 
was taken in the recent work by Ohsuga \& Takahashi (2016) who simulated the evolution
of bolometric intensity assuming Thomson scattering. 
Thomson approximation may be sufficient only for the RMS with negligible fluid 
inertia $\rhof c^2\ll\Urad$. If $\rhof c^2\simgt \Urad$, 
strong bulk Comptonization develops in the RMS and, if treated in the Thomson 
approximation, leads to runaway in photon energy (Blandford \& Payne 1981). 
Scattering with substantial electron recoil, in the Klein-Nishina regime,
becomes inevitable and limits the growth of photon energy.
The electron recoil is also essential in maintaining heat exchange between 
radiation and plasma, even when the photon energies (and the plasma temperature) 
are well below $m_ec^2$.

\subsection{Sample models}

In our two sample simulations the initial flow has dimensionless enthalpy $w=0.1$ and 
magnetization $\sigma=0$ (Model~A) or $\sigma=0.1$ (Model~B). 
The initial average photon energy in the fluid frame is everywhere $3kT\approx 10^{-2}m_ec^2$.
In both simulations we observed how the compressive 
wave with amplitude $\pmax=2$ steepened and formed a pair of shocks at 
time $t\approx L/c$.
We chose a sufficiently large optical depth of the steepening region $\tau_L=\sT nL=20$ 
and observed how the magnetic field and the trapped radiation were advected toward 
$x=0$, building up a strong pressure maximum. This launched the RMS, as described
in \Sect~2, and the two symmetric shocks continued to propagate away from $x=0$.
The amplitude of the shocks $p_0$ slowly grows as they propagate toward the 
asymptotic momentum of the converging flow $\pmax=2$.

%%%%%%%%%%% FIGURE %%%%%%%%%%%%%%%%%%
\begin{figure}[t]
\vspace*{-1.5cm}
\begin{tabular}{c}
\vspace*{-1.9cm}
\includegraphics[width=0.48\textwidth]{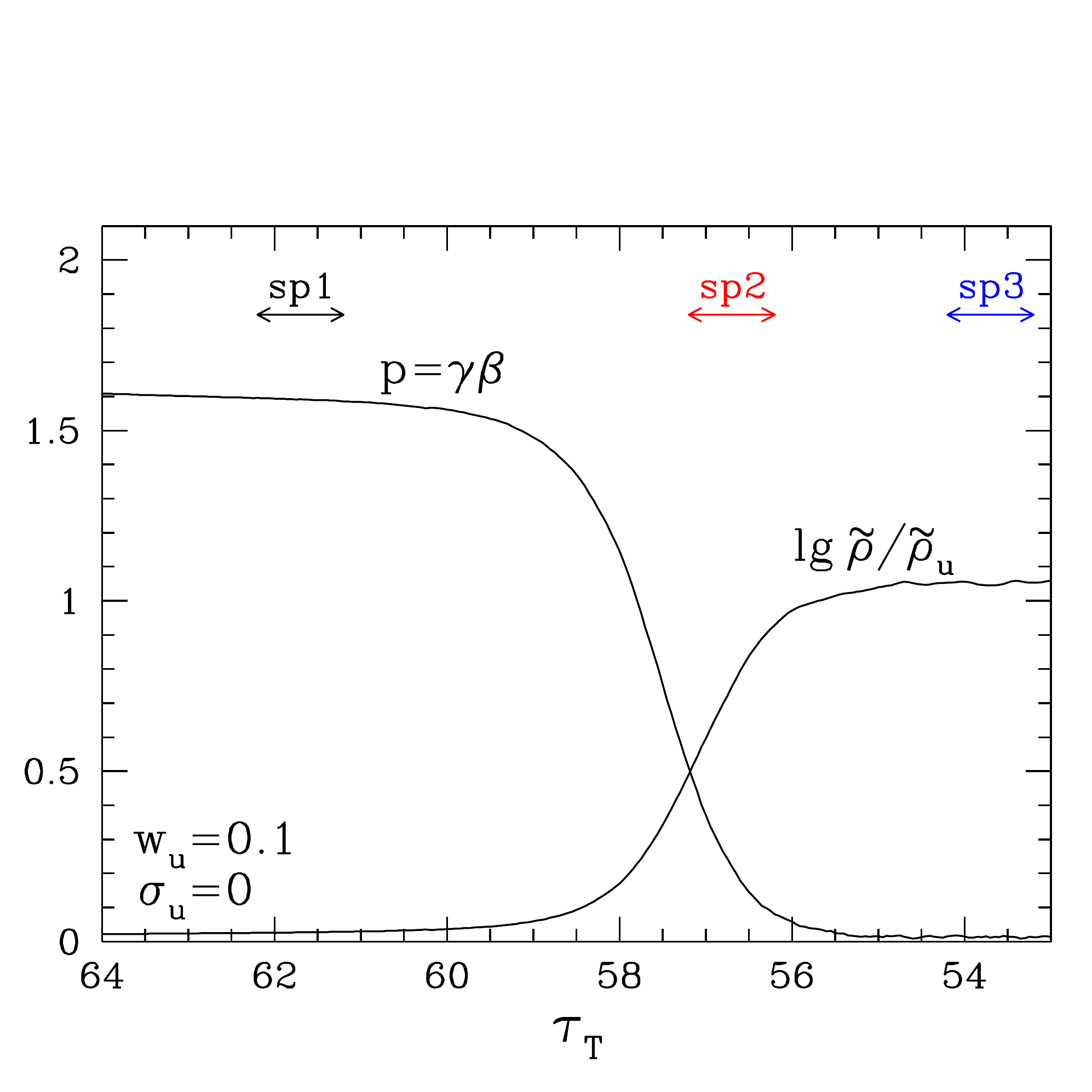}\\
\includegraphics[width=0.48\textwidth]{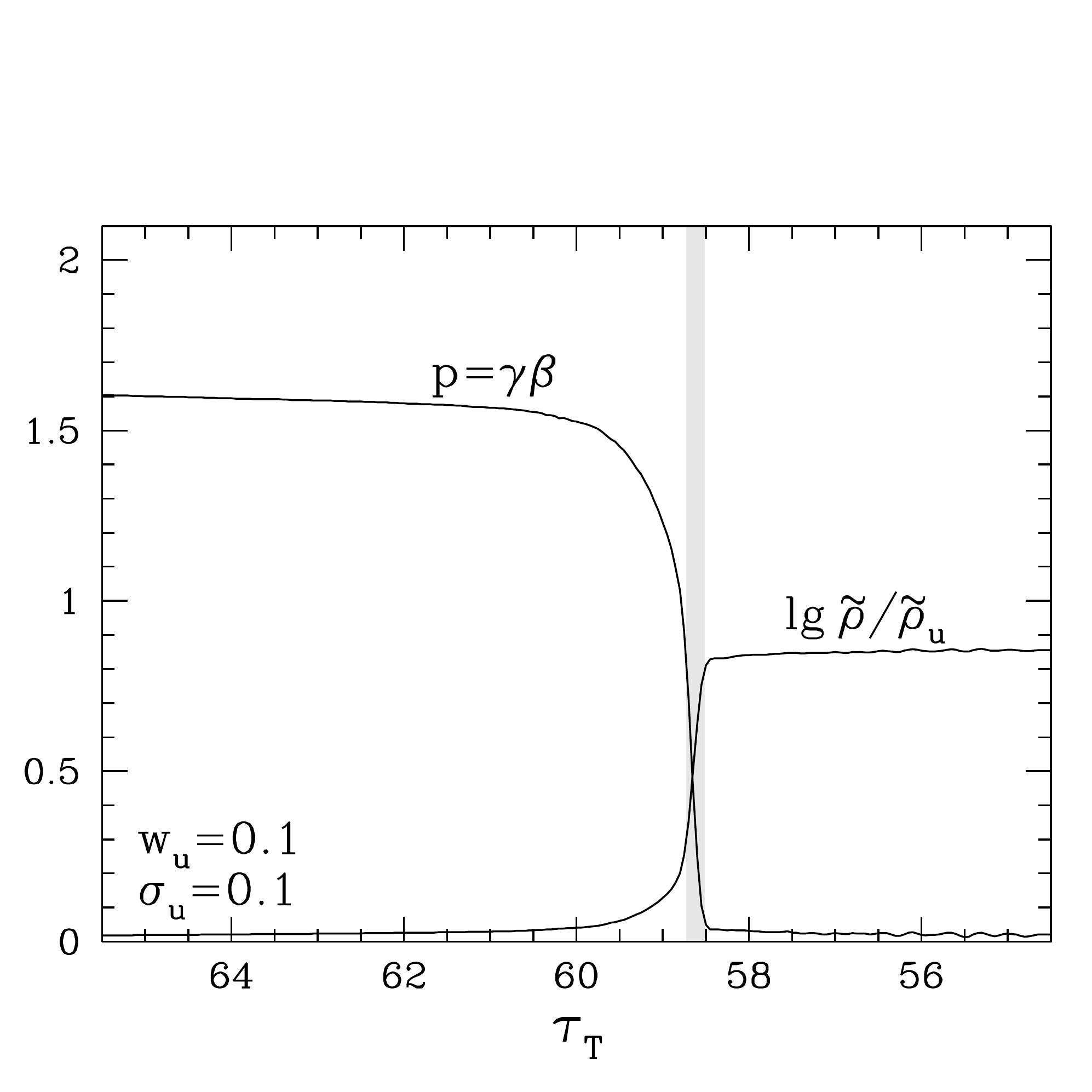}
\end{tabular}
\vspace*{-0.3cm}
\caption{Snapshot of a shock propagating in the flow with upstream radiation
enthalpy $\wu=0.1$ and upstream magnetization $\sigu=0$ (upper panel) 
or 0.1 (lower panel).
The shock is propagating to the left and the Thomson optical depth $\tauT$ is 
measured from the site of shock formation (caustic of the initial supersonic wave).
The solid curves show the profiles of momentum $p=\gamma\beta$ and 
proper density $\rhof$ (normalized to the upstream proper density $\rhofu$).
A strong subshock has formed in the magnetized case; it is highlighted by the grey strip.
The subshock is resolved (not a discontinuous jump) due to a finite viscosity employed in 
the simulation of plasma dynamics; the small subshock thickness 
$\delta\tauT\approx 0.2$ is controlled by viscosity.
Radiation is everywhere simulated directly as a large collection of individual photons 
whose propagation and scattering is followed using the Monte-Carlo technique. 
The photon spectra measured at three locations (indicated as sp1, sp2, sp3)
are shown in Figure~\ref{fig:spectr}.
}
 \label{fig:shock}
 \end{figure}
%%%%%%%%%%% FIGURE %%%%%%%%%%%%%%%%%%

Figure~\ref{fig:shock} shows one of the two symmetric shocks at $t\sim 3L/c$
in Model~A (upper panel) and Model~B (lower panel).
By this time the shock has crossed a Thomson optical depth $\tauT\sim 60$ from 
its formation site, and the upstream momentum has reached $p_0\approx 1.6$.
The shock structure is steady and propagating relative to the downstream with speed 
$v\approx 0.3v_0$. The shock exhibits the jump conditions calculated in \Sect~3. 
In particular, the shock compression ratio in Model~B is $\xi=\rhofd/\rhofu\approx 6.8$.
The jump conditions also give a moderate ratio $\sigd/\wdo\approx 0.6$; it turns 
out sufficient to form a strong collisionless subshock.
The simulation confirms the expectation from \Sect~3 
that the flow magnetization leads to a strong collisionless subshock in the RMS. 
The subshock becomes weak if the magnetization is reduced and disappears
in Model~A with $\sigma=0$.

The observed shock structure in Model~B ($\sigma=0.1$) may be summarized as follows.
The momentum profile 
$p(\tauT)$ is shallow toward the upstream; this part is shaped by radiation 
pressure that gradually decelerates the upstream on a scale comparable to the 
photon mean free path $\lph$. The profile steeply drops toward the downstream and has a
kink connecting to the flat $v\approx 0$. The steep drop is 
accompanied by a narrow and strong spike in the plasma temperature.  
It has a finite thickness in our simulation, 
because we employed a small viscosity to keep shocks resolved by the Lgrangian
grid. The optical depth of the steep drop  $\delta\tauT\approx 0.2$
is sufficiently small to exclude its support by radiation diffusion.
With reduced viscosity, the profile steepens even more and forms a 
discontinuous jump with $\delta\tauT\approx 0$ --- the collisionless subshock.

A large fraction of the subshock energy gets stored in the compressed 
magnetic field, and a fraction is dissipated into plasma heat, as required by the jump 
conditions. This heat adds to 
the RMS entropy generated by photon diffusion between the upstream and downstream.
The viscous heating in the simulation occurs 
near the kink (where the second derivative of the velocity profile peaks) 
and gives a narrow temperature spike immediately behind the velocity jump
--- a stark signature of subshock dissipation.
The high photon-to-electron ratio ($\sim 10^3$ in the simulation) results in 
fast electron cooling, on a timescale $\tIC\ll\lph/c$. It forces the postshock
temperature to quickly return to the Compton equilibrium with the local 
radiation field. Subshock cooling is accompanied by 
additional processes that will be discussed in \Sect~6 below. 

\begin{figure}[t]
\begin{tabular}{c}
\includegraphics[width=0.47\textwidth]{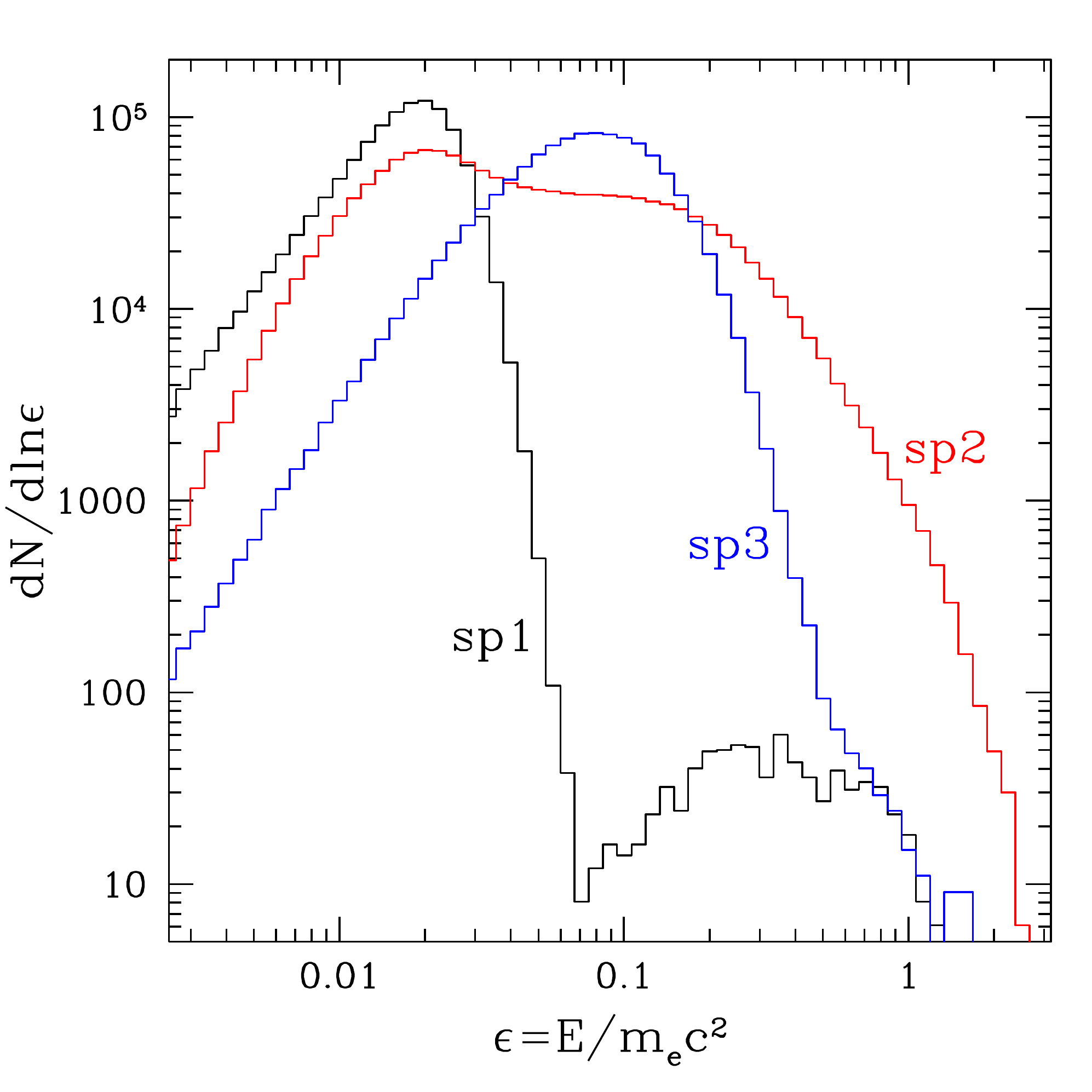}
\end{tabular}
\caption{Photon spectrum $dN/d\ln E$ at three locations in Model~A: upstream (black),
in the middle of the shock (red), and downstream (blue); the three locations are 
indicated in Figure~\ref{fig:shock}. Everywhere photon 
energy $E$ is measured relative to the downstream frame. The overall scale on the 
vertical axis is arbitrary (it was chosen to reflect the photon number per energy bin
used to construct the histogram). The shock has the 
upstream radiation enthalpy $\wu=0.1$, magnetization $\sigma=0$, and amplitude
$p_0=1.6$. The snapshot of the simulation is taken at the same time as in Figure~9. 
The shock structure is nearly steady by this time, and the high-energy 
spectrum inside the RMS approaches saturation. 
The high-energy component seen in the upstream spectrum (the gamma-ray 
precursor) weakens with distance from the shock. The downstream spectrum 
is relaxing toward Compton equilibrium and reaches the Wien shape further 
downstream.
}
 \label{fig:spectr}
 \end{figure}

The Monte-Carlo simulation provides photon statistics that show how the 
radiation spectrum evolves across the shock front. Figure~\ref{fig:spectr} shows this
evolution in the simpler case of $\sigma=0$ where there is no subshock and no
synchrotron emission.
Then the spectrum is shaped by the bulk Comptonization effect, which was 
discussed previously (Blandford \& Payne 1981; Levinson \& Bromberg 2008):
a fraction of photons cross the shock back and forth multiple times, with the energy 
boost $\sim\gamma_0^2$ in every cycle, similar to Fermi diffusive acceleration. 
As a result, the photon spectrum extends somewhat above $m_ec^2$ 
in the fluid frame. Further energy growth is hindered by 
downscattering due to the strong electron recoil 
(and also by photon conversion to $e^\pm$ pairs, see below). At large optical depths 
downstream of the shock, the multiple downscattering of high-energy photons 
drives the spectrum toward a Wien shape in Compton equilibrium with the electrons. 

%%%%%%%%%%% FIGURE %%%%%%%%%%%%%%%%%%
%%%%%%%%%%% FIGURE %%%%%%%%%%%%%%%%%%

\subsection{Pair creation}

Let us consider the weakly magnetized RMS, with no collisionless subshock.
The ability of bulk Comptonization to generate photons with energies 
$E>m_ec^2=511$~keV in the fluid frame implies a significant rate of $e^\pm$ pair 
creation due to reaction $\gamma+\gamma\rightarrow e^++e^-$. 
This reaction was not included in our Monte-Carlo simulations, and below we 
discuss its effect on the RMS structure.

The rate of pair creation can be estimated by noting that the absorbed MeV photons 
are replenished with rate $\sim n_1/t_1$ where $n_1=dn_\gamma/d\ln E$ 
at $E=m_ec^2$ and $t_1\sim (3-10)\lph/c$ is the time it takes a 0.5-MeV photon to 
double its energy through bulk Comptonization.
The radiation spectrum inside the RMS shows 
$n_1\sim 10^{-2} n_\gamma$, where $n_\gamma$ is the total photon density. 
Therefore, one can roughly estimate the pair production rate as
\beq
   \dn_\pm\sim 10^{-3} \,\frac{c\,n_\gamma}{\lph}.
\eeq
Combining this with the characteristic timescale of the shock propagation, $\lph/c$, 
one obtains an order-of-magnitude estimate for the pair density in the RMS,
$n_\pm\sim 10^{-3}n_\gamma$.
The corresponding number of $e^\pm$ created per proton is 
\beq
\label{eq:Z}
   Z_\pm \sim 10^{-3}\,\frac{n_\gamma}{n}=10^2 \left(\frac{n_\gamma/n}{10^5}\right),
\eeq
where $n_\gamma/n\sim 10^5$ is a typical photon-to-baryon ratio in GRBs.

This estimate neglects two effects, which somewhat reduce $Z_\pm$:
\\
(1) \Eq~(\ref{eq:Z}) assumes that the photons reaching $\sim 1$~MeV
are quickly absorbed in $\gamma$-$\gamma$ collisions, neglecting their 
finite free path.
The MeV photons collide with each other, as they are near the threshold for 
$\gamma$-$\gamma$ reaction, $2m_ec^2\approx 1$~MeV. 
Their free path $\lgg$ may be estimated using the cross section for 
$\gamma$-$\gamma$ collision $\sgg\sim 0.1\sT$. Then one finds
$\lgg \sim 10 (\sT n_1)^{-1}\sim 10^3 (Z_\pm n/n_\gamma)\lph$, and 
\beq
\label{eq:load}
   \frac{\lgg}{\lph}\sim \frac{Z_\pm}{10^2}\left(\frac{n_\gamma/n}{10^5}\right)^{-1}\sim 1.
\eeq
Hence a large fraction of $e^\pm$ creation occurs inside the RMS,
as assumed in \Eq~~(\ref{eq:Z}), and so the finite $\lgg$ is not a big change.
\\
(2) \Eq~(\ref{eq:Z}) neglects the effect of $e^\pm$ annihilation.
The annihilation rate $\dnann=(3/8)\sT c\,n_+n_-$ implies a positron lifetime 
$\tann=(8/3)(\sT c\,n_-)^{-1}$. 
The positrons are advected by the flow with a mildly relativistic speed and 
annihilate over a distance that corresponds to Thomson optical depth $\sim 1$, 
comparable to the RMS thickness. Hence annihilation has a moderate reduction 
effect on $Z_\pm$ inside the RMS. 

These estimates show that $Z_\pm$ inside the RMS can approach $\sim 10^2$.
The exact value can be obtained from
detailed numerical simulations and will depend on the upstream temperature,
strength of the shock $\gamma_0$, and $n_\gamma/n$.
$Z_\pm$ peaks inside the RMS and decreases with distance in the downstream,
because of $e^\pm$ annihilation.

The created mildly relativistic pairs 
are almost immediately cooled by inverse Compton (IC) emission and Coulomb collisions 
with the background plasma. The corresponding energy loss rates of an electron with  
Lorentz factor $\gamma_e=(1-\beta_e^2)^{-1/2}$ are given by (e.g. Ginzburg \& Syrovatskii 1964)
\beq
   \dot{E}_{\rm IC}\approx\frac{4}{3}\,\sT\, \Urad \gamma_e^2\beta_e^2,
\eeq
\beq
   \dot{E}_{\rm Coul}\approx\frac{3\ln\Lambda \,\sT n_\pm m_ec^3}{2\beta_e},
\eeq
where $\ln\Lambda=\ln(m_ec^2/\hbar\omega_{\rm pl})\approx 20$ is a Coulomb logarithm
($\omega_{\rm pl}$ is the Langmuir plasma frequency). Their ratio may be written as
\beq
    \frac{\dot{E}_{\rm IC}}{\dot{E}_{\rm Coul}}
    \approx \frac{10^2}{Z_\pm}\,\frac{\Urad}{\rhof c^2}\,\gamma_e^2\beta_e^3,
\eeq
where $Z_\pm=n_\pm/n$ is the number of $e^\pm$ per proton.
For the typical parameters, the Coulomb losses dominate when 
$\gamma_e\beta_e\simlt 1$.
Thus the injected mildly relativistic particle shares a significant fraction of its energy with 
the background plasma and deposits momentum $\sim m_ec$. 
The estimated rate of $\gamma$-$\gamma$ absorption then implies 
the force $f_{\rm abs}\sim \dn_\pm m_ec$ exerted on the plasma.
This additional force is modest compared with the radiation pressure force
$f_{\rm sc}\sim \dn_{\rm sc} \epsp m_ec$, where 
$\dn_{\rm sc}=\sT c\,n_\pm n_\gamma=n_\gamma c/\lph$ is the scattering rate 
and $\eps_p m_ec^2\sim 10$~keV is the average photon energy. Using the 
above estimates one finds 
$f_{\rm abs}/f_{\rm sc}\sim \dn_\pm /\dn_{\rm sc} \epsp\simlt 0.1$.

Pair creation increases the local scattering opacity $\kappa(x)=Z_\pm(x)\sT/ m_p$ 
by the factor $Z_\pm$. In essence, an additional $e^\pm$ ``screen'' is created 
between the upstream and downstream.
The thickness of a relativistic RMS is always a few $\lph$, and $e^\pm$ loading implies 
that the RMS thickness shrinks proportionally to  $\lph=(n_\pm\sT)^{-1}\propto Z_\pm^{-1}$. 
For simplicity, the Monte-Carlo simulations presented in \Sect~5.1 assumed 
$\kappa=const$.
This assumption is not important as long as (1) the shock structure is steady and
(2) the velocity and density profiles of the shock wave are viewed as functions of 
the scattering optical depth (as in Figure~9) rather than the spatial coordinate $x$.
In the plane-parallel approximation (valid when the RMS thickness is much smaller 
than its radius) the actual length $\lph$ does not matter, and the optical depth 
is the natural coordinate.

%###############################################################

\section{Collisionless (sub)shock heating}

In this section, $v_0=\beta_0c$ refers to the strength of the collisionless shock, 
which may be embedded in a stronger RMS.
The shock thickness is microscopic, comparable to the ion Larmor radius.

The heat generated by the shock is initially given to the plasma particles and then
converted to radiation, at some distance downstream of the shock. 
In particular, the post-shock ions receive thermal speeds $v_{\rm th}\sim v_0$,
as the shock thermalizes their upstream bulk speed $v_0$.
Numerical simulations of collisionless electron-ion shocks show that fast collective
processes help the ions to promptly pass a fraction $f_e\sim 0.3-0.5$ of their energy
to the electrons, and both form Maxwellian distributions (Sironi \& Spitkovsky 2011). 

The fraction $f_e$ is less studied for the most interesting shocks in pair-loaded plasma,
with $1\ll Z_\pm\ll m_i/m_e$, where $m_i$ is the ion mass. The existing work on 
pair-loaded shocks with transverse magnetic fields focused on the ultra-relativistic regime,
with application to pulsar wind nebulae. 
It was found that such shocks are capable of positron acceleration through
cyclotron maser instability of gyrating ions
(Hoshino et al. 1992; Amato \& Arons 2006; Stockem et al. 2012). It is unclear 
if particle acceleration may be efficient in moderately relativistic internal shocks in GRBs.

Below we discuss how the postshock plasma radiates its energy, assuming 
a two-temperature state. The electron and ion temperatures immediately behind the shock
are controlled by the parameter $f_e$. The thermal Lorentz factor of the ions, $\gth$,
is related to the upstream Lorentz factor $\gamma_0=(1-\beta_0^2)^{-1/2}$ by
\beq
   \gth-1=(1-f_e)(\gamma_0-1), \qquad {\rm (ions)}.
\eeq
The electrons are heated to much higher Lorentz factors,
\beq
\label{eq:gthe}
  \gamma_{\rm th,e}= f_e(\gamma_0-1)\frac{m_i}{Z_\pm m_e}\gg 1,  
  \qquad {\rm (electrons)},
\eeq 
where $Z_\pm=n_\pm/n$ is the number of $e^\pm$ 
pairs per proton upstream of the collisionless shock.  We will first discuss electron 
cooling and then ion cooling.

\subsection{Electron cooling}

The suddenly heated electrons lose their energy to inverse Compton (IC) scattering on a 
short timescale, 
\beq
\label{eq:tC}
  \tC= \frac{3\,m_ec}{4\,\sT\fKN\Urad\gamma_{\rm th,e}},
\eeq
where the factor $\fKN<1$ describes the Klein-Nishina correction to the Compton 
cooling rate. Even accounting for pair creation, the photon number in GRB jets 
exceeds the electron number by a large factor (3-5 orders of magnitude when the 
jet Lorentz factor saturates). 
Therefore, the electron cooling time $\tC$ is much shorter than the free-path time of 
photons to scattering, $\tC\ll \lph/c$. 
The electrons are also cooled by synchrotron losses on the timescale,
\beq
   t_{\rm syn}=\frac{3m_ec}{4\sT U_B \gamma_{\rm th,e}}.
\eeq
It is shorter than $\tC$ if $U_B>\fKN\Urad$.

The maximum energy of IC photons is determined by $\gamma_{\rm th,e}$,
which depends on $Z_\pm$ (\Eq~\ref{eq:gthe}).
IC photons with energies $\EIC\gg 1$~MeV are processed through $e^\pm$ cascade 
into secondary $e^\pm$ pairs and eventually into photons of energy $\simlt 1$~MeV, 
which are capable of escaping $\gamma$-$\gamma$ absorption 
(Svensson 1987).\footnote{Photons 
      of energy $\EIC$ are absorbed in collisions with photons of energies 
     $E_t\simgt 2m_e^2c^4/E_{\rm IC}$ with cross section $\sgg\sim 0.1\sT$. 
     The mean free path to absorption is $\lgg=(\sgg n_t)^{-1}$ where $n_t$ is the 
     target photon density.
     For instance, 100-MeV photons are absorbed by the Wien peak of the
     radiation spectrum ($E_t\sim 10$~keV in the fluid frame) and their 
     $\lgg\sim 10(\sT n_\gamma)^{-1}$ is tiny. 
      }
As a result, most of $e^\pm$ pairs are created in collisions between photons of energy 
$\EIC\sim 1$~MeV, not much above the $2m_ec^2$ threshold. 
The situation resembles that described in \Sect~5.2.

 The multiplicity of pairs created per shock-heated electron, $\M_\pm$, is maximum 
if synchrotron losses are small $U_B\ll \fKN\Urad$; then it approaches
$\M_\pm\sim 0.2\gamma_{\rm th,e}$. If the magnetic energy is comparable to 
the radiation density, synchrotron losses become dominant.
This significantly reduces the pair yield $Y=\M_\pm/\gthe$ (see Table~1 in Vurm et al. 2011). 

The population of MeV photons with peak density $\nMeV$ spreads from the collisionless
shock over the characteristic distance of their self-destruction in $\gamma\gamma$ collisions,
\beq
  \lgg\sim (\sgg \nMeV)^{-1}.
\eeq
The number of MeV photons emitted per shock-heated electron approximately equals  
the number of pairs they produce, i.e. $\MMeV\sim\M_\pm$.

The $e^\pm$ density in the pair-creation region of size $\sim \lgg$ is somewhat reduced 
by the annihilation reaction with rate $\dnann=(3/8)\sT c n_+n_-$.
The downstream evolution of the positron density $n_+$ is described by the equation,
\beq
\label{eq:ann}
   v_d\,\frac{dn_+}{dx}=\dngg-\frac{3}{8}\,\sT c\,n_+n_-.
\eeq
Here $\dngg\sim\sgg c\,\nMeV^2$ is the pair creation rate, and $\vd\approx v_0/3$ 
is the velocity of the downstream relative to the collisionless 
shock.\footnote{The relation $\vd\approx v_0/3$ holds for plasma-mediated 
      shocks with any amplitude $v_0$, 
      relativistic or non-relativistic, as long as the shock is strong ($v_0$ is much
      greater than the upstream thermal speed). 
      The relation $\vd=v_0/3$ is derived analytically from the jump conditions in 
      the ``monoenergetic gas'' approximation (Beloborodov \& Uhm 2006).}
The advection rate across the pair-creation zone $v_d\,dn_+/dx\sim n_+v_d/\lgg$ 
is comparable to or smaller than the terms on the right side of \Eq~(\ref{eq:ann}). 
Therefore, the pair density is not far from the
annihilation balance, $(3/8)\sT c\,n_+n_-\sim \sgg c\,\nMeV^2$. This is still 
consistent with the crude estimate for the positron density 
$n_+\sim\nMeV$, which determines the Thomson optical depth of the pair 
creation zone, 
\beq
  n_+\sim \nMeV, \qquad \tauT\sim \sT n_\pm\lgg\sim 2,
\eeq

The production of MeV photons that convert to pairs is different from 
the RMS bulk Comptonization discussed in \Sect~5. However, it gives an
optical depth comparable to that of an RMS.

\subsection{Pair breeding upstream of collisionless shocks}

The strong electron heating in collisionless shocks results in the MeV emission
downstream of the shock, however some of the MeV photons can convert to pairs 
in the upstream.
This leads to a special type of self-sustained breeding of  pairs upstream of the shock.

The number of $e^+$ and $e^-$ created per shock-heated electron may be 
written as $\M_\pm=\Mu+\Md$ --- the sum of $e^\pm$ created upstream and 
downstream. Most of the pair-creating MeV photons are emitted 
relatively close to the 
collisionless shock, at a distance smaller than their absorption free-path 
$\lgg$.\footnote{In a GRB jet, the average photon energy in the fluid frame is 
     comparable to 10~keV. Most of the pair-creating IC photons 
     are generated in the Thomson regime, $\EIC\sim \gamma_e^2\times 10{\rm ~keV}$,
     by $e^\pm$ with $\gamma_e\simlt 30$ over length 
    $\lIC\approx v_d\tC\sim 0.1\beta_dm_ec^2/\sT\Urad$, cf. \Eq~(\ref{eq:tC}).
    The length $\lIC$ is much shorter than $\lgg(\EIC)$.
    If pair creation involves a cascade with multiple generations, 
    many of the electrons with 
    $\gamma_e<30$ would be secondary and injected over an extended length
    due to the intermediate steps of $\gamma$-$\gamma$ absorption. 
    However, even in this case the additional steps are relatively short, as
    $\lgg({\rm 100~MeV})\ll\lgg({\rm 10~MeV)}\ll\lgg({\rm 1~MeV)}$.
    }
Therefore, IC photons emitted toward the upstream can overtake the
shock and convert to pairs ahead of it.

A rough estimate
for $\Mu/\M_\pm$ may be obtained assuming that the IC photons are emitted isotropically
in the downstream fluid frame and that photons emitted with angles $\cos\theta>\beta_d$ 
(catching up with the shock) have a sufficiently long $\lgg$ to overtake the shock. Then,
\beq
\label{eq:Mu}
   \Mu\sim \frac{(1-\beta_d)}{2}\M_\pm.
\eeq
A more accurate value for $\Mu$ may be found from Monte-Carlo transfer simulations.
It will be somewhat below the estimate given in \Eq~(\ref{eq:Mu}). 

$\Mu$ particles are injected into the upstream per one electron (or positron) 
heated in the shock. They join the upstream, cool down, and then go 
through the shock like the primary electron did, closing the cycle.
This cycle allows the upstream $e^\pm$ population to reproduce 
itself with the amplification factor $\Mu>1$.

The breeding of upstream pairs is, however, limited, because $\gamma_{\rm th,e}\propto Z_\pm^{-1}$.
The increasing upstream $Z_\pm$ eventually reduces $\gamma_{\rm th,e}$ to $\gcr$
that is marginally sufficient for production of IC photons capable of converting ahead 
of the shock. Then a self-consistent situation is achieved with $\Mu\approx 1$ --- 
the upstream pair population is replenished with no further growth. 
The pair loading factor ahead of the shock is then given by 
\beq
\label{eq:Zpm}
   Z_\pm\approx f_e(\gamma_0-1)\frac{m_p}{\gcr m_e}
     \sim 10^2 f_e \left(\frac{\gcr}{20}\right)^{-1}(\gamma_0-1).
\eeq
The critical value of $\gamma_{\rm th,e}$ can be estimated for a typical spectrum of 
radiation carried by GRB jets, which peaks around 10~keV in the fluid frame.
Then $\gcr\sim 20$ is capable of producing a large number of IC photons with 
$\EIC\simgt 3$~MeV.

The reduction of $\gamma_{\rm th,e}$ to $\gcr$ implies that the downstream $e^\pm$ cascade 
has only one generation and a modest multiplicity $\M_\pm\sim 3$-5. The downstream 
pair loading factor is much higher, $Z_\pm\gg \M_\pm$, due to the breeding of pairs 
ahead of the shock. 

In an RMS, with or without a collisionless subshock, the upstream is decelerated 
by scattering the radiation flux from the downstream. The absorption of MeV photons 
implies an additional deceleration effect.
It involves the upstream deposition of momentum $\sim Z_\pm m_ec$ per ion, which 
may approach $\sim 0.1$ of the ion momentum and is insufficient to significantly 
change the velocity profile of the shock wave. 

This situation is in contrast with 
the previously studied $e^\pm$ creation ahead of {\it external} blast waves  
from GRB explosions (Thompson \& Madau 2000; Beloborodov 2002). 
In that case, the shock is driven by ultra-relativistic ejecta into a low density ambient 
medium of rest-mass $mc^2\sim E_{\rm ej}/\Gamma^2$, where $E_{\rm ej}$ is the 
explosion energy.
The average GRB photon propagating ahead of the shock has energy comparable to 
1~MeV in the rest frame of the ambient medium.
Then the impact of pair creation and radiation pressure on the external medium is huge.
The MeV radiation front is capable of sweeping up the external medium and even clearing 
a vacuum cavity ahead of the ejecta (Beloborodov 2002). Such effects cannot occur in 
internal shocks because they are much weaker and involve a much lower contrast in 
density.

\subsection{Energy runaway?}

Pair breeding described above involves electrons and photons of MeV energies
and does not generate very high-energy particles. 
The most energetic particles have Lorentz factors $\sim\gcr$ immediately 
behind the collisionless shock, and $e^\pm$ pairs injected into the upstream have 
mildly relativistic $\gamma_e\sim 2$. Before reaching the collisionless shock, these
pairs lose energy through Coulomb collisions with the background plasma
and IC cooling (synchrotron cooling of mildly relativistic electrons is 
suppressed by self-absorption).
Compton cooling results from scattering of background photons of energies  
$E_t<100$~keV, and hence the upstream 
produces IC photons with energies $\EIC\sim\gamma_e^2 E_t<m_ec^2$.
Thus, MeV radiation is only generated immediately behind the collisionless shock,
with no amplification cycle in photon energy.

The situation could, however, change if the shock is ultra-relativistic, $\gamma_0\gg 1$.
Then pair breeding occurs with a high energy gain per cycle, tapping into the kinetic 
energy of upstream motion relative to downstream.
It can lead to a runaway cycle of photon energy boosting, which was noticed and described 
as ``converter mechanism'' by Derishev et al. (2003) and ``electromagnetic catastrophe'' 
by Stern (2003). The cycle involves IC emission followed by photon propagation across 
the collisionless shock, $e^\pm$ creation on the other side, and IC emission from 
the created pair back toward the downstream. This effectively implies an 
exchange of IC photons between the upstream and downstream
accompanied by the energy boost $\sim \gamma_0^2$.
The runaway occurs if the cycle is closed with the mean expectation for the final photon 
energy exceeding the initial photon energy.

Consider an initial IC photon of energy $\eps_1 m_ec^2$ in the downstream frame,
and suppose the photon overtakes the shock and converts to an $e^\pm$ pair.
The created particles have 
Lorentz factors $\gamma_e \sim \gamma_0 \eps_1/2$ in the upstream frame, and
cool in the Klein-Nishina regime if $\gamma_e \epsp\simgt 1$ (where 
$\epsp m_ec^2\sim 10$~keV describes the peak of radiation spectrum in the fluid frame).
Then the new IC photons emitted by the $e^\pm$ pair and viewed from the 
downstream frame have energies 
$\eps_2\sim \gamma_0\gamma_e/2\sim (\gamma_0^2/4)\eps_1$.
The cycle is completed when the photon $\eps_2$ propagates into the downstream, 
creates a pair with $\gamma_e\sim\eps_2/2$, and this pair produces IC photons with
$\eps_f\sim \gamma_e/2 \sim (\gamma_0/4)^2\eps_1$.
This rough estimate suggests that a runaway cycle, $\eps_f>\eps_1$,
requires $\gamma_0>4$. It also requires two other conditions:
\\
(1) The electron cooling free path $\lIC$ should not exceed the free path of photons
it produces, $\lgg$; otherwise the IC photon is absorbed before it has a chance to 
cross the shock. This condition is not satisfied for high-energy IC photons, 
$\eps\gg\epsp^{-1}$, if the target radiation spectrum at $\eps_t<\epsp$ has the 
photon index $\alpha=d\ln n_t/d\ln \eps_t<0$.
If $\alpha>0$, IC scattering and $\gamma$-$\gamma$ absorption of photons 
$\eps>\epsp^{-1}$ involve the same main 
targets $\eps_t\sim \epsp$  
with comparable cross sections $\sgg\sim\sigma_{\rm IC}\sim (\eps\epsp)^{-1}\sT$; 
then $\lIC\sim\lgg$.
\\
(2) The magnetic fields should be weak; otherwise synchrotron losses become increasingly 
dominant over IC emission in the deep Klein-Nishina regime and suppress the production
of high-energy IC photons.

In summary, the runaway of photons with growing energies 
requires an ultra-relativistic collisionless shock and weak magnetic fields. 
This combination is unlikely at large optical depths where the formation of 
strong collisionless shocks requires strong magnetic fields.

\subsection{Ion cooling}

The energy fraction kept by the ions behind the collisionless shock, $1-f_e$, is not 
easily radiated --- an ion cannot directly radiate its energy, because of its large mass 
$m_i$. The ions gradually lose their energy through Coulomb collisions with $e^\pm$ or 
through nuclear collisions. Both processes are relatively slow, and the ion cooling 
can be a bottleneck for postshock heat conversion to radiation.

The $e^\pm$ plasma behind the shock quickly becomes much colder than the ions. 
Frequent Compton scattering enforces kinetic equilibrium of electrons 
(and positrons) with local radiation at the Compton temperature $k\TC\sim 10-50$~keV, 
which corresponds to a thermal speed around $0.3c$. 
In the first approximation, the hot, mildly relativistic ions behind the collisionless shock
view electrons as targets at rest, and the ion cooling timescale due to Coulomb 
collisions is approximately given by (e.g. Ginzburg \& Syrovatskii 1964),
\beq
   \tCoul=\frac{(\gth-1)m_ic^2}{\dot{E}_{\rm Coul}}=
   \frac{2\bth(\gth-1)m_i}{3\ln\Lambda \,\sT n_\pm m_ec},
\eeq
where $n_\pm$ is the local density of electrons and positrons and 
$\ln\Lambda\approx 20$ is the Coulomb logarithm.

During time $\tCoul$, the hot ions are advected through the distance
\beq
  \lCoul=\vd\,\tCoul=\frac{v_0}{3}\, \tCoul,
\eeq
where $\vd\approx v_0/3$ is the velocity of the downstream relative to the 
collisionless jump. The Thomson optical depth of the ion cooling region is 
\beq
  \tauT=\sT n_\pm \lCoul
  =\frac{2m_i\,\beta_0\,\bth(\gth-1)}{9m_e\,\ln\Lambda}.
\eeq

A significant fraction of radiation produced by the ion cooling at distance $\lCoul$ 
behind the shock can diffuse back into the upstream if $\tauT v_d/c\simlt 1$. 
Otherwise, radiation is trapped and advected away from the shock, missing the 
chance to affect the upstream velocity profile. We conclude that the ion heat 
produced by the collisionless jump is effectively lost for the RMS if
\beq
   \frac{2\,m_i\,\beta_0^2\,\bth(\gth-1)}{27\,m_e\,\ln\Lambda}>1.
\eeq 
This condition is satisfied for collisionless jumps with amplitude
$p_0=\gamma_0\beta_0\simgt 1$. In this case, the delay in ion cooling tends to 
reduce the radiative precursor and increase the amplitude of the collisionless jump.

The mildly relativistic ion (proton) temperature behind a strong subshock,
$kT_i\sim 1$~GeV, leads to inelastic nuclear collisions between the protons, 
producing pions. Half of the pion energy is lost to neutrino emission while the other half 
is converted to $e^\pm$ with Lorentz factors $\gamma_e\approx m_\pi/m_e\sim 300$.
This injection of relativistic pairs by p-p collisions sustains an $e^\pm$ cascade 
in the downstream region of thickness $\lCoul$. 
The cascade is similar to that  triggered by inelastic n-p 
or n-n collisions in a neutron-loaded jet (Derishev et al. 1999; Beloborodov 2010). 

The timescale for inelastic p-p collisions
is $t_{pp}\approx (cn_p\sigma_{\rm inel})^{-1}$ where 
$\sigma_{\rm inel}=f_{\rm inel}\sigma_n$ is a substantial fraction of the nuclear cross 
section $\sigma_n\approx \sT/20$. The ion heat lost to inelastic collisions before it is 
given to $e^\pm$ via Coulomb collisions is determined by the ratio 
\beq
  \frac{\tCoul}{t_{pp}}=\frac{2\bth(\gth-1)m_i\sigma_{\rm inel}}{3\ln\Lambda \, Z_\pm m_e\sT}
  \sim \frac{1}{Z_\pm}.
\eeq
One can see that $e^\pm$ loading reduces the role of p-p collisions behind the shock.

%###############################################################

\section{Transformation of shocks near photosphere}

As an RMS emerges from the photosphere of the outflow, it must transform into
a pure collisionless shock: radiation becomes decoupled from the plasma and 
the shock must be sustained by collective (collisionless) plasma processes.
This transformation occurs through the growth of the collisionless subshock inside 
the RMS.

A key feature described in \Sects~5.2 and 6 is that the shocks are dressed in 
$e^\pm$ pairs. This delays their transition to transparency and thus delays the 
transition to a pure collisionless state. While the upstream 
and far downstream regions are already transparent to radiation, the shock 
itself remains opaque until its pair dress becomes optically thin. The estimated
pair-loading factor $Z_\pm$ inside the shock front is comparable to $10^2$, 
and hence the effective photospheric radius should be increased by a factor $\sim 10^2$. 
The shock ``carries'' the photosphere with it and keeps radiating right 
at the photosphere rather than crosses it. 

Pair creation is strong even before the shock approaches the photosphere.
A weakly magnetized relativistic shock generates pairs through bulk Comptonization, 
as described in \Sect~5.2. In the magnetized case, pairs are generated by the 
collisionless subshock, as described in \Sect~6. In any case, when the shock 
attempts to emerge from the photosphere and become a pure collisionless jump, 
pair creation with a large $Z_\pm$ is inevitable.

Details of the photospheric transition depend on the shock amplitude 
$p_0=\gamma_0\beta_0$ and magnetization $\sigma$.
The radius $\Rph^\pm$ where the shock eventually becomes transparent 
satisfies the condition $\Rph^\pm\sim Z_\pm\Rph$ where $\Rph$ is the
photosphere in the absence of pair creation. Note that our estimates for $Z_\pm$  
assumed that most MeV photons ($\sim\Gamma\times$MeV in the lab frame)
emitted by the shock at a radius 
$r$ convert to pairs with a free path $\ll r$ in the lab frame, which corresponds to 
$\lgg\ll r/\Gamma$ in the fluid frame. Taking into account that $\sgg\sim 0.1\sT$,
this requires a sufficiently large ``compactness'' parameter,
\beq
\label{eq:l}
   l=\frac{\Urad}{m_ec^2}\,\sT\frac{r}{\Gamma}\gg 10,
\eeq
where $\Urad=L_{\rm rad}/4\pi r^2 c\Gamma^2$ is the radiation energy density in the 
fluid frame and $L_{\rm rad}$ is the isotropic equivalent of the observed GRB luminosity.
The compactness parameter is related to the characteristic 
Thomson optical depth of the outflow $\tauT=n_\pm\sT r/\Gamma$,
\beq
\label{eq:l1}
   l\sim \frac{n_\gamma}{Z_\pm n}\,\frac{\bar{E}}{\Gamma m_ec^2}\,\tauT
    \sim \frac{10^3}{Z_\pm}\,\tauT,
\eeq
where $\bar{E}\sim 1$~MeV is the average photon energy 
measured the static lab frame, and $n_\gamma/n\sim 10^5$ is the typical photon-to-baryon
ratio (we use the numerical values typical for GRBs). 
If $l\simlt 10-30$, the $\Gamma\times$MeV photons do not convert to pairs; 
only photons of significantly higher energies are quickly absorbed. 
A rough estimate for $\Rph^\pm$ may be obtained by combining \Eqs~(\ref{eq:Zpm}) and 
(\ref{eq:l1}), which gives $\Rph^\pm\sim 30\Rph$. Its exact value 
depends on the parameters of the explosion. 

A moderate magnetization of the flow $\sigma\sim 0.1$ implies 
that the collisionless shock expanding from $\Rph$ to $\Rph^\pm$ is a strong source 
of synchrotron photons. Unlike the IC photons (many of which convert to pairs) the 
synchrotron photons are soft 
and dominate the low-energy part of the shock radiation spectrum. 
Overall, the photospheric radiation released by the $e^\pm$-dressed shock should have 
a broad nonthermal spectrum around $\bar{E}\sim 1$~MeV in the lab frame. It extends 
from the synchrotron self-absorption energy (well below 1~MeV) to the GeV break 
shaped by $\gamma$-$\gamma$ absorption. 

Detailed calculations of the emitted spectrum are deferred to a future paper.
Here we note that the structure and observational appearance of the shock 
emerging from $\Rph$ is affected by the development of 
radiation anisotropy in the fluid frame. In any relativistic outflow, radiation 
develops a strong forward beaming in the fluid frame at optical depths $\tauT\simlt 10$
(Beloborodov 2011). As the $e^\pm$-dressed shock expands from $\Rph$ to $\Rph^\pm$,
its radiation maintains a strong beaming.
This effect breaks the symmetry between shocks propagating forward and backward 
relative to the plasma outflow (shocks form in pairs propagating
in the opposite directions relative to the fluid, see \Sect~2). The downstream of a backward internal shock 
is radially ahead of the upstream; hence the forward beaming reduces the efficiency of 
radiation diffusion from downstream to the upstream.
This accelerates the development of the collisionless shock 
and also influences the effective optical depth $\tauT$ of the $e^\pm$ dress. 
In contrast, for a forward-propagating shock the downstream is radially behind the 
upstream. Then beaming assists sending photons into the upstream with a decreasing 
angle relative to the radial direction.

%############################################################

\section{Nuclear collisional dissipation}

\subsection{Neutral particles in sub-photospheric shocks}

In general, shock formation occurs when the steepening of the compressive wave 
is  stopped by momentum (or heat) transfer due to the finite mean free path of particles. 
In a multi-component fluid, the importance of different 
components for the shock structure is determined by their contributions to viscosity 
and thermal conductivity. Both are controlled by the diffusion coefficient,
\beq
    D=\frac{1}{3}\,\ell\,\bar{v},
\eeq
where $\ell$ is the mean free path and $\bar{v}$ is the characteristic thermal speed
of a given species of particles.
In particular, the viscosity coefficient created by each species may be estimated as 
\beq
   \mu=D\xi \frac{H}{c^2},
\eeq
where $\xi$ is the fractional contribution of the species to the relativistic 
enthalpy density of the flow $H$, which includes the rest-mass energy 
($H/c^2$ serves as the effective inertial mass density in relativistic hydrodynamics).

Consider now shock formation deep below the photosphere.
A simple comparison of viscosity coefficients of different components of the 
preshock flow allows one to judge their importance for the shock structure.
In a first approximation, electrons, positrons, ions, and magnetic fields, move together 
as a single, strongly coupled fluid. In contrast, neutral particles --- photons and especially 
neutrons --- have large free paths and a large diffusion coefficient which leads to 
significant transfer of momentum and energy between the upstream and downstream, 
shaping the profile of the shock wave.

The consideration of diffusion coefficients assumes that the particles are not 
decoupled from the flow, i.e. their mean free-path time $\ell/\bar{v}$ is smaller than the 
age of the flow (measured in its rest frame).
Neutrinos may carry a significant fraction of the flow energy in GRBs, however they 
escape freely and do not contribute to viscosity. Photons contribute to viscosity 
at radii $r\simlt\Rph$ and neutrons --- at radii $r\simlt R_n$ where 
$R_n\approx \Rph(\sigma_n/Z_\pm\sT)$ is the neutron decoupling radius.

\subsection{Neutron-mediated shock wave}

GRB jets are expected to carry a significant number of free neutrons 
(Derishev et al. 1999; Beloborodov 2003). 
Their $\beta$-decay is delayed proportionally to the outflow Lorentz factor $\Gamma$
and occurs at a characteristic radius $R_\beta\approx 8\times 10^{15}(\Gamma/300)$~cm, 
which is much larger than the typical photospheric radius $\Rph\sim 10^{12}-10^{13}$~cm.
Neutrons are coupled to the plasma 
through nuclear collisions with cross section $\sigma_n\approx 3\times 10^{-26}$~cm$^2$,
which is 20 times smaller than Thomson cross section.
Neutrons have the largest mean free path, carry a significant fraction of the flow 
momentum, and hence can dominate the flow viscosity, affecting the formation of shocks.
Neutron migration across the shock assist the momentum exchange between the 
upstream and downstream on a scale comparable to the neutron mean free path $\ell_n$,
shaping a broad, ``neutron-mediated'' shock wave.

\begin{figure}[t]
\begin{tabular}{c}
\hspace*{-0.7cm}
\includegraphics[width=0.53\textwidth]{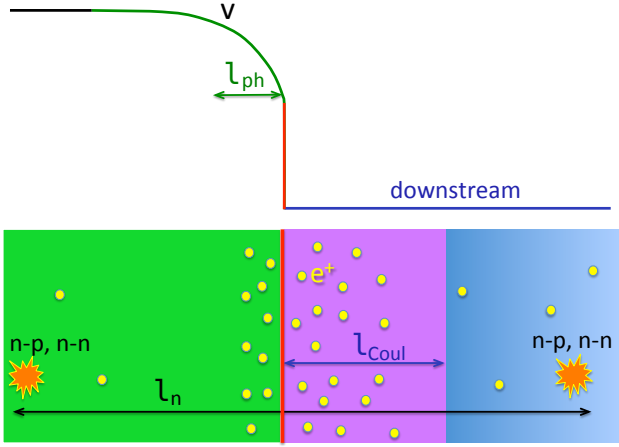}
\end{tabular}
\caption{Structure of a relativistic internal shock in a hot, opaque, magnetized outflow.
The green part of the velocity profile is shaped by the radiation pressure on a scale
comparable to the photon mean free path to scattering $\lph$. The vertical red part is 
the collisionless subshock, which heats ions to a mildly relativistic temperature and 
electrons to an ultra-relativistic temperature (\Sect~6). 
The ultra-relativistic electrons are quickly cooled, producing synchrotron and 
IC photons, which create $e^\pm$ pairs ahead and behind the shock.
The ion cooling occurs on a longer scale $\lCoul$, which may exceed the RMS 
thickness if the subshock is relativistic. 
Some of the hot ions experience inelastic p-p collisions before cooling.
In the presence of a free neutron component in the flow, the shock wave is partially 
shaped by neutron migration between the upstream and downstream.
The mean neutron free path $\lln$ is the longest scale in the shock structure. 
The migrating neutrons are stopped by nuclear collisions. Inelastic nuclear collisions 
result in injection of $e^\pm$ with Lorentz factors $\gamma_e\sim m_\pi/m_e\sim 300$, 
distributed over a broad region $\sim \ell_n$. The nuclear collisions also emit neutrinos
with observed energies $\sim \Gamma m_\pi c^2$ where $\Gamma$ is the outflow 
Lorentz factor.
}
 \label{fig:scheme}
 \end{figure}

This wave can have a strong subshock mediated by radiation,
as the neutron collisions alone are unable to stop everywhere the velocity profile from 
steepening.
If the neutron fraction of the outflow momentum is modest, 
the main momentum jump in the wave occurs in the subshock.
Then the wave is better described as an RMS with a neutron precursor 
rather than a neutron-mediated shock.
The RMS itself can have a strong collisionless subshock, depending on the flow magnetization. The resulting wave structure
is schematically shown in Figure~\ref{fig:scheme}.

\subsection{Nuclear collisions around RMS}

Let us first consider small radii where the outflow rest-mass energy 
is dominated by radiation, $\Urad\gg\rhof c^2$. Then the energy of neutrons 
(including their rest mass) is small compared with the energy dissipated in the RMS. 
The RMS thickness $\lsh$ is comparable to the photon mean 
free path $\lph$. It is much smaller than the neutron mean free path $\ell_n$, 
\beq
    \frac{\ell_n}{\lph}\sim Z_\pm\,\frac{\sT}{\sigma_n}\approx 20 Z_\pm,
\eeq
where $Z_\pm=n_\pm/n\geq 1$ determines the reduction of the photon free path 
in the plasma enriched by $e^\pm$ pairs.

Neutrons brought by the upstream view the RMS as a discontinuity in the fluid velocity 
profile. They cross it ballistically and dissipate their energy $(\gamma_0-1)m_nc^2$ 
and momentum $p_0m_nc$ at the characteristic distance $\sim \ell_n$ downstream 
of the shock. Their collisions with the downstream nuclear matter
create a relativistically ``hot'' neutron (and ion) component embedded in the photon gas. 
Some of the hot neutrons propagate back into the upstream and collide there. 
They form a precursor of the shock --- a ``neutron pillow'' that somewhat 
decelerates the upstream and thus reduces the strength of the RMS.

%%%%%%%%%%% FIGURE %%%%%%%%%%%%%%%%%%
%%%%%%%%%%% FIGURE %%%%%%%%%%%%%%%%%%

As long as neutron rest-mass density $\rhof_nc^2$ is small compared with 
the enthalpy of the upstream photon gas, $4P_0$, the effect of neutron transport on the 
RMS amplitude is small.
For example, consider a radiation-dominated jet with an asymptotic (saturation) 
Lorentz factor $\Gamma_{\rm sat}$
at a small radius where the Lorentz factor is $\Gamma=0.2\Gamma_{\rm sat}$. 
At this early stage, 80\% of the jet energy is carried by radiation, and about 20\% 
is carried by baryons (neglecting here the energy of the magnetic field and $e^\pm$).
Assuming that half of baryons are neutrons, one can estimate 
$\rhof_nc^2/4P_0\sim 0.1$, so roughly 10\% of the shock energy is dissipated by 
nuclear collisions in this example. 
This dissipation occurs in the region of thickness $\sim \ell_n$ around the RMS.
Even the modest amount of collisional dissipation at small radii (large optical 
depths) is important because it efficiently generates photons, as discussed below.

As the expanding and accelerating outflow experiences adiabatic cooling,
 the ratio $\rhof_nc^2/4P_0$ increases and 
the role of collisional dissipation grows. When the Lorentz factor $\Gamma$
approaches its maximum value $\Gamma_{\rm sat}$, i.e. when radiation density 
$\Urad$ becomes comparable to $\rhof c^2$,
a large fraction of the shock energy becomes dissipated through nuclear collisions. 
The wave front of thickness $\sim \ell_n$ --- the neutron-mediated shock wave ---
propagates as long as the flow is opaque 
to nuclear collisions. When the wave approaches the neutron decoupling radius $R_n$ 
the subshock (RMS) grows, and at larger radii $r>R_n$ the fraction of energy 
dissipation through nuclear collisions is reduced  as $R_n/r$.

The  RMS itself is not capable of producing ultra-relativistic particles, and so
collisional dissipation around the RMS plays a key role in this respect. 
Even in moderately relativistic shocks the neutron collisions are energetic enough to 
be inelastic. Such collisions produce mildly relativistic pions which quickly 
decay into ultra-relativistic $e^\pm$ with Lorentz factors $\gamma_e\sim m_\pi/m_e$
(Derishev et al. 1999).
This generates an inverse Compton cascade which produces copious $e^\pm$ pairs.
The resulting pair loading factor $Z_\pm=n_\pm/n$  is comparable to 10 as long as 
the jet magnetization parameter is below 0.1 (Beloborodov 2010; Vurm et al. 2011). 
Far downstream of the shock, where the baryons have cooled, the inelastic collisions
and the cascade end; here the pairs annihilate, if they still have time to do so before 
freezing out. The freeze-out happens when the outflow expansion timescale 
becomes shorter than the annihilation timescale; this occurs
when the flow approaches the photosphere (Beloborodov 2010).

Note that neutrons can convert to protons (and protons to neutrons) in 
inelastic nuclear collisions. This enables the ``converter'' mechanism
for baryon acceleration proposed by Derishev et al. (2003).
Numerical results of Kashiyama et al. (2013) suggest 
that the converter mechanism becomes efficient for 
ultra-relativistic shocks, $\gamma_0\simgt 4$.

%############################################################

\section{Discussion}

This paper focused on sub-photospheric internal shocks in relativistic explosions,
their dissipation mechanism and structure. One question of observational interest 
is whether the shocks are capable of producing ultra-relativistic 
electrons. Energetic electrons at large optical depths
emit synchrotron radiation (without self-absorption) 
and thus boost the photon number carried by the flow, which is later released at 
the photosphere. Another interesting question is how the shock evolves and radiates
as it approaches the photosphere.

As long as the flow is opaque and radiation dominates its energy density, 
photon transport plays a leading role in shaping the shock front.
Its thickness is then comparable to the photon mean free path. 
The radiation-mediated shock (RMS) is not capable of 
electron acceleration by the standard Fermi mechanism, since the electron 
radiates its energy faster than it can cross the shock. Photons 
experience significant energy gains by crossing the shock back and forth multiple times. 
This ``bulk Comptonization'' upscatters photons up to the MeV band (in the fluid frame);
further upscattering is hindered by the energy loss due to electron recoil in scattering. 
The photon upscattering beyond $\sim 1$~MeV is also stopped by the absorption 
reaction $\gamma+\gamma\rightarrow e^-+e^+$.
The produced mildly relativistic $e^\pm$ pairs immediately cool down due to 
fast Coulomb and IC losses.

All this would suggest that sub-photospheric shocks are inefficient in producing 
particles with energies $E\gg m_ec^2$ in the fluid frame. However, a more realistic 
shock picture significantly differs from the simple RMS, 
in particular when one takes into account that the outflow 
carries magnetic fields and free neutrons (see Figure~\ref{fig:scheme}).
The shock wave is capable of generating ultra-relativistic electrons in two ways: 
\\
(1) A strong collisionless subshock forms in the RMS. We have shown that this is 
inevitable (even deep below the photosphere) if the flow is sufficiently magnetized. 
A mildly relativistic collisionless subshock heats the electrons to 
an ultra-relativistic temperature $T_e$. Their IC emission 
breeds $e^\pm$ pairs in the upstream and regulates the shock structure to a 
self-consistent state with the postshock temperature $kT_e\sim 10 m_ec^2$ (\Sect~6.2). 
This temperature is high enough to 
generate interesting synchrotron radiation without self-absorption, which   
is strong for electrons with Lorentz factors $\gamma_e\simlt 10$ (Vurm et al. 2011). 
The subshock fails to generate particles with $\gamma_e>10$ if it is weak (the flow 
is weakly magnetized) while pair loading remains strong, $Z_\pm\sim 10^2$, due to 
the conversion of bulk-Comptonized MeV photons (\Sect~5.2).
\\
(2) Inelastic nuclear collisions inject $e^\pm$ pairs with Lorentz factors 
$\sim m_\pi/m_e\sim 300$ in the fluid frame. This mechanism 
becomes particularly efficient if the outflow carries free neutrons, as they can migrate 
across the RMS, making the shock wave partially mediated by neutrons 
(Figure~\ref{fig:scheme}).

The synchrotron losses of energetic electrons generated by either mechanism 
imply significant photon production. The synchrotron photons may carry a small 
fraction of radiation energy, however, their {\em number} is significant. 
GRB jets tend to experience photon-starvation in the ``Wien zone'' at optical depths 
$10^2<\tauT<10^5$ (Beloborodov 2013). In this zone, the heated photons have a Wien 
(rather than Planck) spectrum, i.e. a Bose-Einstein distribution with a 
non-zero chemical potential. The production of low-energy photons is followed by their 
quick Comptonization to the Wien peak. The addition of photons shifts
the peak to lower energies, as the energy per photon is reduced.
This effect regulates the observed peak position of the GRB spectrum
that is eventually released at the photosphere (Vurm \& Beloborodov 2016).
Synchrotron photons produced outside the Wien zone, i.e. at smaller optical depths 
$\tauT<10^2$, are only partially Comptonized toward the Wien peak and form the 
low-energy part of the prompt GRB spectrum with the characteristic photon index 
$\alpha\sim 1$ (see Vurm \& Beloborodov 2016; Thompson \& Gill 2014).

Nuclear collisions in the shock front also generate neutrinos.
Neutrino emission from migrating and colliding neutrons in GRB jets 
was previously discussed in some detail
(Derishev et al. 1999; Bahcall \& M\'esz\'aros 2000; M\'esz\'aros \& Rees 2000a).
The typical energy of neutrinos produced by this mechanism is 
$\sim\Gamma m_\pi c^2\simgt 10$~GeV
and they are detectable by IceCube (Bartos et al. 2013; Murase et al. 2013).
Sub-photospheric internal shocks were also proposed to emit 
ultra-high-energy neutrinos (M\'esz\'aros \& Waxman 2001; Razzaque et al. 2003).
This proposal assumed efficient ion acceleration by the Fermi diffusive mechanism.
This mechanism does not operate in an RMS.
Diffusive acceleration is also suppressed in the collisionless subshock with a 
transverse magnetic field, which advects the particles downstream before they 
have a chance to cross the shock many times (Sironi \& Spitkovsky 2011).
An oblique magnetic field could help this process to occur.

Detailed studies of sub-photospheric shock structure require numerical simulations.
Previous work focused on the search of a steady-state solution 
for the RMS (e.g. Levinson \& Bromberg 2008; Budnik et al. 2010; Tolstov et al. 2015), 
which may be found by iterations.
Instead, we suggest two techniques that permit direct time-dependent 
simulations of shock formation, as demonstrated in \Sects~4 and 5.
Our simulations are set up to follow the evolution of 
an internal compressive wave, which leads to formation of a pair of shocks and 
their subsequent quasi-steady propagation.
The shock structure is obtained from first principles, by simulating the 
time-dependent radiative transfer in the moving plasma.

We have implemented two methods for such simulations: (1) solving the  
 radiative transfer equation coupled to the flow dynamics and (2) tracing 
individual photons and their interaction with the moving plasma using 
Monte-Carlo technique. Both methods 
show the structure of the established shock wave (Figures~8 and 9) and 
reproduce the jump conditions described in \Sect~3. 
The simulations verified the formation of a strong subshock in magnetized RMS.
In particular, in GRB jets, a moderate magnetization $\sim 0.1$ is sufficient to 
generate a strong subshock. 

A curious feature indicated in Figure~\ref{fig:scheme} and discussed in \Sect~6 
is the delayed cooling of the ions heated in the collisionless subshock. 
If the subshock is relativistic, the ion cooling length $\lCoul$ can exceed the RMS 
thickness. When the ion
energy is finally radiated, the produced radiation is trapped and advected downstream,
missing the chance to diffuse upstream and affect the shock velocity profile.
The delayed ion cooling also implies that some ions experience inelastic nuclear collisions 
and emit neutrinos even in the absence of a free neutron component.

We have estimated the pair-loading factor $Z_\pm \sim 10^2$ in the RMS with or
without a collisionless subshock. 
The $e^\pm$ pairs are produced in collisions between MeV photons, which are 
generated by two mechanisms. RMS without a strong collisionless subshock 
produces MeV photons only through bulk Comptonization in a relatively cold converging
flow, which is close to Compton equilibrium with local radiation at $kT\ll m_ec^2$. 
In the presence of a collisionless jump, MeV photons are produced by IC cooling of 
$e^\pm$ heated in the jump to $kT_e\sim 10 m_ec^2$.

As the GRB jet expands from the central engine 
the structure of internal shocks and dissipation mechanism change. 
As long as the jet energy is dominated by radiation,
$\Urad\gg\rhof c^2$ and $\Urad\gg U_B$, the shock structure is described by a 
unique solution, which has no collisionless subshock (Figure~8). 
Bulk Comptonization in such deep sub-photospheric shocks is also suppressed, 
and their structure is conveniently described by the ``force-free'' 
radiative transfer solution for the bolometric intensity.
This regime can occur at small radii in GRB jets, where 
the outflow Lorentz factor $\Gamma\ll\Gamma_{\rm sat}$.

Nuclear dissipation due to neutron migration across the RMS increases with $\Gamma$ 
and approaches its maximum near the radius of Lorentz factor saturation. 
It remains high until the neutron decoupling radius $R_n$, and then it declines as 
$(r/R_n)^{-1}$.

The shock ``breakout''  at the photosphere occurs through the growth of the 
collisionless subshock in the RMS until radiation completely decouples from the 
plasma. Eventually the entire velocity jump becomes mediated by collective 
plasma processes, regardless of magnetization.
This somewhat resembles the shock breakout in non-relativistic supernova explosions 
(Waxman \& Loeb 2001;  Giacinti \& Bell 2015). 
However, there is a special feature: $e^\pm$ pair 
creation near the collisionless shock sustains an optical depth 
$\tauT\simgt 1$ even after the background electron-ion plasma becomes transparent.
Effectively, the shock carries the photosphere with it until it expands by an additional
factor $\sim 30$, continually producing photospheric emission. This emission will be 
observed as a prominent pulse of nonthermal radiation in the GRB light curve.

In the observer frame, the emission from the $e^\pm$-dressed shock extends up to the 
GeV band, where $\gamma$-$\gamma$ absorption shapes a break in the spectrum. 
The high-energy photospheric pulses overlap, in observer time, with the GeV flash 
that is produced by the $e^\pm$-loaded external blast wave at a larger 
radius $R\sim 10^{16}$~cm  (Beloborodov et al. 2014; Hasco\"et et al. 2015). 
These pulses may explain the observed variable component of GeV emission 
superimposed on the smooth GeV flash at early times (Ackermann et al. 2013).

Internal shocks are a particularly promising heating mechanism for jets that are not 
magnetically dominated.
Comparison of detailed models of photospheric radiation with observed GRB spectra 
suggests a moderate magnetization in the sub-photospheric region, 
$\sigma\sim 10^{-2}-10^{-1}$ (Vurm \& Beloborodov 2016). This does not however 
exclude a stronger magnetization close to the central engine, allowing for magnetic
dissipation that reduces $\sigma$ as the jet expands. 
In this scenario, the early heating would be
dominated by magnetic dissipation. Even in this regime shocks can occur 
and dissipate significant energy, as follows from the jump conditions discussed in \Sect~3.

\vspace{0.2in}
I thank Hirotaka Ito, Christoffer Lundman, and Indrek Vurm for discussions and
comments on the manuscript. This work was supported by 
NSF grant AST-1412485, NASA grant NNX15AE26G,
and a grant from the Simons Foundation (\#446228, Andrei Beloborodov)

%###############################################################

\begin{appendix}

\section{Location of shock formation in a supersonic compressive wave}

A supersonic wave converging toward 
the caustic $x=0$ forms a pair of shocks at the Lagrangian coordinates $\pm\xx$.
The value of $\xx$ can be estimated as follows.

Consider the streamlines 
that experience smooth compressive deceleration, due to the conversion of kinetic energy 
to enthalpy. For a streamline with a given Lagrangian coordinate $x_0$ the characteristic 
time $\td(x_0)$ and location $\xd(x_0)$ where deceleration occurs is given by 
\Eqs~(\ref{eq:td}) and (\ref{eq:xd}).
Next, note that the smooth compressive deceleration to a subsonic speed 
is possible as long as the density of the gas accumulated in the 
subsonic region, $\rhod$, is comparable to the density of the ballistic flow approaching it,
$$
  \rhob\approx \frac{\rho_0}{1+\vp\,\td}.
$$
Since $\rho\propto P^{1/\adind}$ and pressure in the subsonic region is not far from 
uniform (it tends to equilibrate on the sound crossing time), $\rhod$ is roughly uniform,
\beq
   \rhod \approx \frac{x_0}{\xd}\,\rho_0.
\eeq 
The density ratio is given by
\beq
\label{eq:f}
    f(x_0)\equiv\frac{\rhob}{\rhod}\approx\frac{1+v_0\,\td/x_0}{1+\vp \td}.
\eeq
Using \Eq~(\ref{eq:td}), one can exclude $\td$ and obtain 
\beq
\label{eq:f1}
   f(x_0)\approx 1-\left(\frac{v_0}{\vp x_0}-1\right)
        \left[\left(\frac{(\alpha-1)v_0^2}{2c_0^2}\right)^{1/(\adind-1)}-1\right].
\eeq
The unity in the last term (in square brackets) may be neglected for streamlines
with $v_0^2\gg c_0^2$.
One can see that $f<1$ and $f$ is close to unity for small $|x_0|$. It sharply drops 
when $|x_0|$ exceeds a characteristic $\xx$, and formally even changes sign.
The characteristic $\xx$ may be estimated from the condition $f\sim 0$.

For waves with amplitudes $v_{\rm max}\gg c_0$ one finds that $\xx$ is
much smaller than the wavelength. Therefore, in the calculation of $\xx$ one can use 
the Taylor expansion of the velocity profile around $x_0=0$,
\beq
\label{eq:Taylor}
    v_0(x_0)=-a\,x_0+\frac{b}{6}\,x_0^3 +{\cal O}(x_0^4), 
\eeq
where $a=-\vp(0)$ and $b=v_0^{\prime\prime\prime}(0)$. Here we took into account that 
the second derivative $v_0^{\prime\prime}$ vanishes at $x_0=0$ (recall that we chose 
the zero $x$-coordinate at the location of the caustic where  
$-v_0^\prime$ is maximum, see \Sect~2.1). A linear expansion 
$v_0=-ax_0+{\cal O}(x_0^3)$ would not be sufficient
--- a uniform $\vp(x_0)$ would imply a uniform compression of the ballistic flow, 
with no pressure gradient that could cause deceleration. One can also see from 
\Eq~(\ref{eq:f1}) that it is the deviation from the linear velocity profile 
$v_0/\vp x_0-1\approx bx_0^2/3a$ that controls the drop of $f$ below unity, and
hence controls shock formation.

For the supersonic streamlines with $v_0^2\gg c_0^2$ \Eq~(\ref{eq:f1}) yields the relation
\beq
     1-f\approx \frac{b}{3a}\,x_0^2\left[\frac{(\alpha-1)a^2x_0^2}{2c_0^2}\right]^{1/(\adind-1)},
\eeq
and hence
\beq
\label{eq:xx}
     {\xx}^{2\adind}\approx \frac{3^{\adind-1}}{\adind-1}\,2c_0^2\, a^{\adind-3}\,b^{1-\adind}.
\eeq
The coefficients $a$ and $b$ in the Taylor expansion (\ref{eq:Taylor}) can be estimated 
as $a\approx\pmax c/L$ and $b\approx\pmax c/L^3$ for a smooth initial profile of the wave; 
these relations are exact for a sine profile $p_0(x_0)=-\pmax\sin(x_0/L)$.
Substitution to \Eq~(\ref{eq:xx}) gives
\beq
\label{eq:xx2}
   \frac{\xx}{L}\approx \chi \left(\frac{c_0}{c\pmax}\right)^{1/\adind},
\eeq
where $\chi\approx 3^{(\adind-1)/2\adind}[2/(\adind-1)]^{1/2\adind}\sim 2$
for the relevant range of $4/3<\adind<2$.
Numerical simulations (similar to the sample model shown in Figure~3, with 
different $\adind$ and $c_0$) 
provide the accurate location of shock formation and confirm the 
scaling predicted by \Eq~(\ref{eq:xx2}), with a slightly larger numerical coefficient
$\chi\approx 3-4$.

\end{appendix}

%###############################################################

\end{document}